\documentclass[preprint,authoryear,12pt]{elsarticle}

\usepackage{graphicx}

\usepackage{amssymb}

\usepackage{bm}
\usepackage{amsmath}
\usepackage{tabularx}
\usepackage{colortbl} 
\usepackage{url} 

\journal{Journal of Japanese and International Economy}

\begin{document}

\begin{frontmatter}



\title{What Causes Business Cycles?\\
Analysis of the Japanese Industrial Production Data}


\author{Hiroshi Iyetomi}

\author{Yasuhiro Nakayama}

\address{Department of Physics, Niigata University, Niigata 950-2181, Japan}

\author{Hiroshi Yoshikawa}

\address{Faculty of Economics, University of Tokyo, Tokyo 113-0033, Japan}

\author{Hideaki Aoyama}

\address{Department of Physics, Kyoto University, Kyoto 606-8501, Japan}

\author{Yoshi Fujiwara}

\address{ATR Laboratories, Kyoto 619-0288, Japan}

\author{Yuichi Ikeda}

\address{Hitachi Ltd, Hitachi Research Laboratory, Ibaraki 319-1221, Japan}

\author{Wataru Souma}

\address{College of Science and Technology, Nihon University, Chiba 274-8501, Japan}

\begin{abstract}
We explore what causes business cycles by analyzing the Japanese industrial production data. The methods are spectral analysis and factor analysis. Using the random matrix theory, we show that two largest eigenvalues are significant. Taking advantage of the information revealed by disaggregated data, we identify the first dominant factor as the aggregate demand, and the second factor as inventory adjustment. They cannot be reasonably interpreted as technological shocks. 
We also demonstrate that in terms of two dominant factors, shipments lead production by four months. Furthermore, out-of-sample test demonstrates that the model holds up even under the 2008-09 recession. Because a fall of output during 2008-09 was caused by an exogenous drop in exports, it provides another justification for identifying the first dominant factor as the aggregate demand. 
All the findings suggest that the major cause of business cycles is real demand shocks.
\end{abstract}


\end{frontmatter}

\newcommand\cc{\bm{C}}
\newcommand\ta{\widetilde{a}}
\newcommand\tw{\widetilde{w}}
\newcommand\ga{_{\alpha,g}}
\newcommand\tn{N'}

\section{Introduction}\label{sec:intro}

What causes business cycles? 
A glance at \citet{haberler1964prosperity} reveals that all kinds of theories 
had already been advanced by the end of the 1950's. 
In recent years, macroeconomists had been so confident of their 
skillful policy management as to hail the ``Great Moderation". 
\citet{lucas2003macroeconomic} in his Presidential Address to the meeting of the American Economic Association made the following declaration: 

\begin{quote}
Macroeconomics was born as a distinct field in the 1940's, as a part of the intellectual response to the Great Depression. The term then referred to the body of knowledge and expertise that we hoped would prevent the recurrence of that economic disaster. My thesis in this lecture is that macroeconomics in this original sense has succeeded: Its central problem of depression prevention has been solved, for all practical purposes, and has in fact been solved for many decades 
\citep[p.1]{lucas2003macroeconomic}.
\end{quote}

Ironically, when Lucas delivered this address, the American economy had been on the steady road to the post-war worst recession, perhaps even depression.
Business cycles are still with us, and await further investigation.

The real business cycle (RBC) theory pioneered by \citet{kydland1982time}, arguably the most dominant school today, regards changes of total factor productivity as the primary cause of business cycles. This theory is basically the neoclassical equilibrium theory, and minimizes the role of the government's stabilization policy. The Keynesian theory, in contrast, attributes business cycles to fluctuations of real aggregate demand \citep{summers1986some, mankiw1989real, tobin1993price}, and assigns the government and the central bank the substantial role of stabilization. The fundamental problem, namely what causes business cycles, remains unresolved. 

The standard approach to study this problem is to explore the relative importance of technology shocks on one hand and demand shocks on the other as the sources of business cycles by making various identifying restrictions on macro time-series models: See \citet{shapiro1988sbc}, \citet{blanchard1989dea}, \citet{francis2005trb} and works cited therein. Though variables used and methods differ in details, they all share the same key identifying restriction. Namely, the level of output (or real GDP) is determined in the long-run only by supply shocks such as technology shocks. Conversely, the long-run effect of demand shocks on output is assumed to be nil. On this assumption, not directly observable disturbances are sorted into technology shocks and demand shocks.$^1$

A majority of works routinely make this assumption without questioning its validity. However, we maintain that the assumption is not actually so tenable as is commonly believed. The problem is that in the study of business cycles, technology shocks are usually taken exogenous as typified by real business cycle theory \citep{kydland1982time}. Here, we must observe that modern micro-founded macroeconomics is curiously inconsistent in that while new growth theory puts so much emphasis on the endogeneity of technical progress, business cycle theory is naively content with taking advances in technology as exogenous. New technology or technical progress is certainly not a free lunch! First of all, it must be generated by R\&D investment. Once new technology is found out, it can augment production only by way of investment which embodies such new technology. Now, investment including R\&D investment, is, of course, much affected by cyclical fluctuations. Monetary factors as well as real factors affect investment. Thus, to the extent that new technology is embodied in investment, demand shocks are bound to affect the pace of technical progress. Therefore, we can not naively accept the standard identifying restriction that demand shocks do not leave the permanent effects on the level of output.

One way to proceed without this questionable assumption is to take advantage of disaggregated data. The representative works on business cycles cited earlier all use the macro variables. However, business cycles as the conspicuous macro phenomena are characterized by the fact that output movements across broadly defined sectors and types of goods move together \citep{mitchell1951happens}. In Mitchell's terminology, they exhibit high conformity. It is then most natural to seek possibly unobserved common factors underlining sectoral comovements as the basic factors causing business cycles. Moreover, the separate effects of such unobserved factors on production, shipments and inventory are expected to provide us with valuable information on the causes of business cycles.

Factor models precisely fit out purpose
(\citet{sargent1977business}, 
\citet{stock1998diffusion, stock2002forecasting}). 
\citet{stock1998diffusion, stock2002forecasting} based on this method develop the Diffusion Index to be used for improving forecasting the cyclical fluctuations. 
In this paper, we take the same approach for a different purpose. By examining the relationship between production, shipments, and inventory of twenty-one types of goods, we explore what causes business cycles. 
We also show that the model estimated based on the pre-Lehman crisis 
(September 15th, 2008) can reasonably explain the post-Lehman  (2008-09) 
recession of the Japanese economy.
Our findings are in favor of the old Keynesian view \citep{tobin1993price}. 

\section{Indices of Industrial Production}\label{sec:data}
We use the monthly data of the 
Indices of Industrial Production (IIP) compiled by the Ministry of Economy, 
Trade and Industry (METI) of Japan.
They are indices of industrial production, producer's shipments, and 
producer's inventory of finished goods. 
Details of IIP are in the web site of \citet{iip}.
Here, we simply note that the indices are compiled based on nearly 500 items 
(496 for production and shipment indices; 358 for inventory).
We use the data for the 240 months from January 1988 to December 2007.
(The data itself is available starting on January 1977, but some items are missing
before January 1988.)
We use IIP classified by types of goods, which is listed in Table~\ref{tab:21}.

\def\ga{_{\alpha,g}}
\def\tn{N'}

We denote the IIP by $S\ga(t_j)$,
where $\alpha=1,2,$ and $3$ denotes production (value added), shipments, 
and inventory, respectively.
Also, $g=1,2, \dots, 21$ denotes 21 categories of goods shown in Table \ref{tab:21}, and 
$t_j=j \Delta t$ with $\Delta t=1$ month, and $j=1,2, \dots, N$
for $N$  $(=240)$ months starting in 1/1988 and ending in 12/2007.
A typical $S_{\alpha,1}(t_j)$ is shown in Figure \ref{fig:data_g1new}.

The logarithmic growth-rate $r\ga(t_j)$ 
is defined by the following:
\begin{equation}
r\ga(t_j):=\log_{10}\!\left[ \frac{S\ga(t_{j+1})}{S\ga(t_j)}\right],
\label{rdef1}
\end{equation}
where $j$ runs from 1 to $N-1=239\ (:=\tn)$.
The time-series $r_{\alpha,1}(t_j)$ is plotted in the bottom panel of Figure~\ref{fig:data_g1new}.

The normalized growth rate $w\ga$ is then defined by 
\begin{equation}
w\ga(t_j):= \frac{r\ga(t_j)-\langle r\ga\rangle_t}{\sigma\ga}\,
\label{wnormal}
\end{equation}
where $\langle \cdot \rangle_t$ denotes average over time $t_1, \dots, t_{\tn}$
and $\sigma\ga$ is the standard deviation of $r\ga$ over time.
Thus the set $\{w\ga(t_1), w\ga(t_2), \dots , w\ga(t_{\tn})\}$
has the average equal to zero, and the standard deviation equal to one.
The normalization is done because the normalized data are to be used
in the subsequent analysis of random matrix theory.
We also note that the information revealed by
correlation rather than covariance is sufficient for
our study of lead-lag relationships among production, shipments and inventories.
In the next section, we will examine the cyclical behavior
of the normalized growth rate (\ref{wnormal}).

\section{The Power Spectrum}
\def\tw{\widetilde{w}}
There are two basic problems in the study of business cycles.
First is to identify the periodicity.
Secondly, we must identify common factors causing sectoral comovements. 

In this section, we identify the periodicity. Some economists
are skeptical of the stable periodicity of business cycles 
\citep{diebold1990nonparametric}.
It is certainly true that when we examine the {\it sample path property} of stochastic process,
we face difficulties in finding or even clearly defining the periodicity. However, it
does not necessarily mean that we cannot usefully explore the periodicity of 
business cycles. We resort to the standard spectral analysis (see \citet{granger1966tss}, for example).

The Fourier decomposition is done as follows:
\begin{equation}
w\ga(t_j)=\frac1{\sqrt{\tn}}\sum_{k=0}^{\tn-1}\tw\ga(\omega_k) e^{-i \omega_k t_j},
\label{fd}
\end{equation}
where the Fourier frequency $\omega_k :=2\pi k/(\tn\Delta t)$, 
so that $\omega_k t_j =2\pi kj/\tn$.
Because $w\ga(t_j)$ are real-valued, we have$^2$,
\begin{equation}
\tw\ga^*(\omega_k)=\tw\ga(\omega_{\tn-k}).
\label{wrel}
\end{equation}
The normality of $w\ga(t_j)$ implies that 
\begin{equation}
\tw\ga(0)=0, \quad \sum_{k=0}^{\tn-1}\left|\tw\ga(\omega_k)\right|^2=\tn.
\label{cnorm}
\end{equation}
We define the averaged power spectrum $p(\omega_k)$  as follows:
\begin{equation}
p(\omega_k)=\frac{1}{M}\sum_{\alpha=1}^3 \sum_{g=1}^{21} 
\left|\tw\ga(\omega_k)\right|^2,
\label{defps}
\end{equation}
where $M=63$, the total number of the time series.
Recall that we have 21 categories of goods each for which there are three different variables,
namely production, shipments, and inventory. We note that the following relation is satisfied:
\begin{equation}
\sum_{k=0}^{\tn-1}p(\omega_k)=\tn,
\label{tcons}
\end{equation}
due to Eq.~(\ref{cnorm}).

Figure~\ref{fig:powerspectrum} shows the behavior of $p(\omega)$
of the seasonally-adjusted data$^3$ and the non-adjusted original data
as a function of the period $T:=2\pi/\omega$ [months].
Hereafter, we will use units of months for $T$ unless otherwise noted.
Since the Fourier components of $j=1, ..., (N'-1)/2$
are actually identical to the components of $j=N'-1, ... , (N'+1)/2$ because of Eq.~(\ref{wrel}), 
we plot only the former half.
Therefore, the lower limit of the data in this plot is $2*N'/(N'-1)\approx 2.008$.
Also, the non-adjusted data is multiplied by a factor 7 in this plot 
so that the comparison becomes easier. 

The original data has high peaks$^4$ at $T=$ 1 year, 6 months, and 3 months 
showing the presence of significant seasonal fluctuations. 
We observe that they are well-eliminated in the adjusted data.$^5$
We will use the seasonally-adjusted data for the following analysis.

We observe the significant power at low frequencies,
$T=60$ months (5 years) and $T=40$ months (3 years and 6 months).
One might be tempted to argue that strengths of the cycles $T=120$ 
and $T=239$ are comparable to them.
However, for the 239 months our data covers, the $T=120$ oscillation occurs
only twice, and the $T=239$ only once. 
Therefore, we disregard them as ``one-time" events.
Also, one might argue that $T=24$ appears to be a possible cycle.
But so are many peaks for $T<24$.
In general, peaks for shorter periods tend to be the results of
noise, and one needs to discard them. The criteria for 
selection of significant periods will be discussed in 
relation to dominant eigenmodes in Section 5.

We can demonstrate the robustness of the $T=60$ and $T=40$ cycles. 
These two cycles have the wavenumber (number of periods in the whole region of 239 months)
$k=4$ and $k=6$ respectively, 
so that more accurate values of the periods are $N'/k\approx 59.75$ 
and $N'/k\approx 39.83$.
Therefore, if the number of observations, $N'$ is different from the present value, there is actually no Fourier-coefficient at these periods. Because of this, one may cast a doubt on the robustness of the two cycles. In what follows, we will examine this point to demonstrate that the two cycles are quite robust.

Specifically, we have done the following two analyses.
\begin{description}
\item[{\rm Discrete Fourier transformation for chopped data:}]
We take the first $S$ months and carry out the discrete 
Fourier transformation as in the manner of Eq.~(\ref{defps}).
\item[{\rm Continuous Fourier transformation:}]
We allow $\omega$ to take any value in Eq.~(\ref{defps}), and obtain the 
power-spectrum as a continuous function of $T=2\pi/\omega$.
This is an over-complete analysis, but should be sensitive
to existence of periods away from the discrete values allowed
in the original analysis.
\end{description}

The results are shown in Figure~\ref{fig:robust}.
The dots connected with thick lines are the same discrete 
power-spectrum as in Figure~\ref{fig:powerspectrum}
while the small dots connected by thin lines are the discrete Fourier transform
of the chopped data for $S=234, 229, \cdots, 164$.
The solid curve with gray shades shows the result of the continuous Fourier transformation 
on which the original power-spectrum data-points lie.
We find that these results are basically same.
For the short periods, say less than $T<24$, all the analyses give
essentially identical results, and for the longer periods,
the discrete spectrum of the chopped data shows behavior similar to 
the averages of the continuous analysis.

In Figure \ref{fig:robust}, we observe three cycles A, B, and C 
for the continuous plot, and also for some of the chopped data.
The peak A of the continuous plot is at the period $T\approx 36.96$
and the peak B is at the period $T\approx 59.62$. These are both
{\it very} close to the peaks we have found by our discrete
analysis. This confirms the robustness of our basic finding.

The third peak C is at $T\approx 100.16$. The periods allowed in
our original power-spectrum in Eq.~(\ref{defps}) are
for $T=N'/2\approx 119.5$ and $T=N'/3\approx 79.67$, both of which are
in the outskirts of this peak as is seen in this plot.
However, the $T\simeq100$ cycle is also a 
``one-time" event; it only occurred two and a half times
during our observation of 239 months.

We conclude that the $T=40$ and $T=60$ cycles are significant and
robust.
The peaks and troughs of Japan's business cycles are officially determined by the Cabinet Office (prior to 2000, the Economic Planning Agency). In the post-war period (October, 1951 -- January, 2002), the Japanese economy had experienced twelve cycles. The average length of twelve cycles is 50.3 months with the maximum 83 months and the minimum 31 months. The periodicity we have identified, $T=40$ and $T=60$ months is, therefore, quite reasonable.

\section{The Dominant Factors Identified by the Random Matrix Theory}\label{sec:rmt}
\def\cc{\bm{C}}
Next, we must identify the significant common factors underlying sectoral 
comovements. 
By the common factors, we mean a small number of factors that explain the
correlation of the original data as much as possible.
We resort to the principal component analysis.

First, we obtain the correlation matrix $\cc$ as follows;
\begin{equation}
C_{\alpha,g;\beta,\ell}:=\langle w_{\alpha,g} w_{\beta,\ell}\rangle_t
\label{Cdef}
\end{equation}
whose diagonal elements are 1 by definition of the normalized growth rate $w\ga$.
Because $\alpha$ and $\beta$ run from 1 to 3 and $g$ and $\ell$ run
from 1 to 21, the matrix $\cc$ has $M\times M$ ($M=63$) components.

To uncover the unobservable common factors, we examine the eigenvalues $\lambda^{(n)}$
and the corresponding eigenvectors $\bm{V}^{(n)}$ of the matrix $\bm{C}$;
\begin{equation}
\bm{C}\,\bm{V}^{(n)} = \lambda^{(n)} \bm{V}^{(n)}.
\end{equation}
The eigenvalues $\lambda^{(n)}$ satisfy the trace-constraint;
\begin{equation}
\sum_{n=1}^{M}\lambda^{(n)}=M.
\label{tracec}
\end{equation}
The correlation matrix $\cc$ is written in terms of the eigenvalues and eigenvectors as follows:
\begin{equation}
\bm{C}=\sum_{n=1}^{M}\lambda^{(n)}\bm{V}^{(n)}\bm{V}^{(n){\rm T}}
\label{CVV}
\end{equation}

The distribution of the eigenvalues $\lambda^{(n)}$
is plotted in Figure~\ref{fig:eigenvalue_distribution}.
We see that there are a few isolated large eigenvalues, and that the rest concentrates in the smaller 
region. The large eigenvalues and associated eigenvectors correspond to the dominant factors 
which explain comovements of the fluctuations of different goods.

The eigenvector corresponding to the largest eigenvalue (which we call ``the first eigenvector" hereafter) is the first principal component (PC) that accounts for as much of the correlation as possible in the $M$-dimensional space. The eigenvector for the second largest eigenvalue (``the second eigenvector") is the second principal component that accounts for as much of the correlation as possible in the subspace orthogonal to the first eigenvector, and so on.

In the principal component analysis (PCA), it remains an unresolved problem
how to choose significant factors. To solve this problem,
many procedures and guidelines have been proposed \citep{jackson1991user, jolliffe2002principal, peres2005many}.
For example, the Kaiser-Guttman rule \citep{guttman1954some,kaiser1960application}
which is very popular in practice suggests that only factors
having eigenvalues greater than one are to be retained.
The scree graph named by \citet{cattell1966scree} is also widely used. The scree graph is 
a kind of rank size plot of the eigenvalues of a correlation matrix. In this graph, the abscissa is the rank and the ordinate is the eigenvalue.
The typical shape of the scree graph is a downward-sloping curve, and the slope changes from
the ``steep'' section to the ``shallow'' section. Accordingly, we can obtain PCs by looking for the point that separates the ``steep'' and the ``shallow''.
Although the scree graph is used mainly in PCA, Cattell proposes the same method for the factor analysis as well.

The problem of testing the significance of a few
dominant factors is also a challenge to exploratory factor analysis (FA), or the
so-called approximate dynamic factor model such as 
\citet{stock1998diffusion, stock2002forecasting}.
Determining the number of unknown dominant factors has been, in fact, studied in the rapidly growing literature. 
\citet{bai2002dnf}, for example, propose a criteria for the selection of dominant factors in large dimensions of $N$ and $T$.
They basically use the information criteria in a framework of model selection. Their analyses are, however, done on the asymptotic assumption of large dimensional panels.

\bigskip\noindent
{\bf Random Matrix Theory}
\medskip

Though how to determine the number of significant components is still an open issue in PCA and FA, the parallel analysis (PA) proposed by \citet{horn1965rationale} is considered very likely to be the best procedure \citep{zwick1986comparison}. In PA, the scree graph derived from the actual correlation matrix is compared with that from the
random matrix that parallels the original data set.$^6$
Because the multivariate time-series of IIP contain uncorrelated shocks and
``measurement'' errors, the empirical correlation matrix is to some extent random. 
It is, therefore, reasonable to compare the properties of the empirical correlation matrix $\bm{C}$ to a {\it null hypothesis} on purely random matrix which one would obtain with time-series of totally uncorrelated fluctuations. This is the basic idea of the PA.

We apply the method of random matrix theory (RMT) to the problem of
comparing the properties of the empirical correlation matrix $\bm{C}$
with the null hypothesis based on the purely random matrix.
The RMT has made great success in physical sciences, and has quickly
found applications to various disciplines. We note that the RMT has been
applied to finance for a decade (see the textbook by
\citet{bouchaud2000} and references therein) and macroeconomics
\citep{ormerod2008rmt}.

PA is widely applied to various disciplines, but it requires
the Monte-Carlo simulation to obtain the scree graph for the random matrix. The drawback of the Monte-Carlo simulation is cost of computation time. Several
methods have been proposed to overcome this problem. For
example, the tables of simulated eigenvalues of the random matrix with different sample size and number of variables are provided by \citet{buja1992remarks}. However, if we
can obtain the theoretical formula for the probability density function (PDF) for the eigenvalue of the random matrix with different sample size and number of variables,
we can overcome this drawback of the Monte-Carlo simulation
and PA. That is what we will do in what follows.

Let us consider an ensemble of random matrices
in which all the elements of a square matrix $\bm{A}$ of size $M\times M$
are iid (independent and identically-distributed)
random variables with the constraint that $A_{ij}=A_{ji}$. As
a consequence of the central limit theorem, in the limit of large
matrices, the distribution of eigenvalues has universal properties,
which are mostly independent of the distribution of the elements
$A_{ij}$. In particular, it can be shown that if the $A_{ij}$'s are
iid random variables of zero mean and variance $\langle A_{ij}^2
\rangle=\sigma^2/M$, the probability density function (PDF) of
eigenvalues of $\bm{A}$ is given by the so-called
``semi-circle law'' \citep{wigner1955characteristic}:
\begin{equation}
  \rho(\lambda_{\text{A}})=\frac{1}{2\pi\sigma^2}
  \sqrt{4\sigma^2-\lambda_{\text{A}}^2},
\label{rmt_semicirc}
\end{equation}
for $|\lambda_{\text{A}}|\leq 2\sigma$ and zero elsewhere (see
\citet{bouchaud2000}).

Its relevance to our study of correlation is obvious when sample
correlation matrix is written as $\bm{C}=\bm{A}^{\text{T}}\bm{A}$
where $\bm{A}$ is a rectangular matrix of size $M\times N$ ($M$
and $N$ respectively the number and length of time-series), and
$\bm{A}^{\text{T}}$ is the transpose of $\bm{A}$. If $M=N$, the
eigenvalue $\lambda$ of $\bm{C}$ can be obtained from the
obvious relation $\lambda=\lambda_{\text{A}}^2$. If it is assumed that the
elements of $\bm{A}$ are iid random variables, the PDF of
eigenvalues of $\bm{C}$ can be easily calculated from the semi-circle
law (\ref{rmt_semicirc}) by using the relation
$\rho(\lambda)d\lambda=2\rho(\lambda_{\text{A}})d\lambda_{\text{A}}$
(for $\lambda_{\text{A}}>0$) as
\begin{equation}
  \rho(\lambda)=\frac{1}{2\pi\sigma^2}
  \sqrt{\frac{4\sigma^2-\lambda}{\lambda}},
\label{rmt_semicirc2}
\end{equation}
for $0\leq\lambda\leq 4\sigma^2$ and zero elsewhere.

If $M\neq N$, one can derive a similar PDF
(\citet{marcenko1967distribution}; see \citet{bouchaud2000} for a
simple derivation). Using our notations in
this paper of $M$ and $\tn$, the PDF for the
eigenvalue of the $M\times M$ correlation matrix $\bm{C}$ for $M$ stochastic
variable of length $\tn$ with standard deviation $\sigma$ is given by
\begin{equation}
\rho(\lambda)=
\begin{cases}\displaystyle
\frac{Q}{2\pi\sigma^2}
\frac{\sqrt{(\lambda_+-\lambda)(\lambda-\lambda_-)}}{\lambda}
& \mbox{for } \lambda_-<\lambda<\lambda_+, \\[10pt]
0 & \mbox{otherwise,}
\end{cases}
\label{rmte1}
\end{equation}
in the limit of 
\begin{equation}
M\rightarrow\infty, \quad \tn\rightarrow\infty
\label{limits}
\end{equation} 
with $Q:=\tn/M$ fixed.
The bounds $\lambda_\pm$ for $\lambda$ are defined as follows:
\begin{equation}
\lambda_\pm:=\sigma^2\frac{(1\pm\sqrt{Q})^2}{Q}.
\label{rmte2}
\end{equation}
Note that (\ref{rmte1}) is reduced to (\ref{rmt_semicirc2}) when $M=\tn$.

Though the above results (Eqs.(\ref{rmte1})--(\ref{rmte2})) hold
exactly true only asymptotically in the limit (\ref{limits}),
we emphasize that the RMT is actually robust with respect to
sample size; {\it it holds up not only asymptotically but also for finite
sample with nontrivial autocorrelation\/}. We demonstrate it in Appendix A with
the help of a Monte-Carlo analysis of randomized data. Given this result,
in what follows, we perform significance test of
eigenvalues based on the RMT advanced in physics.

The dashed line in Figure~\ref{fig:eigenvalue_distribution} shows the prediction of RMT, 
Eq.~(\ref{rmte1}) and (\ref{rmte2}), where in the present case,
\begin{equation}
\sigma=1, \quad
Q=\frac{239}{63}\approx 3.79, \quad
\lambda_+\approx 2.29, \quad
\lambda_-\approx 0.24.
\label{rmtbound}
\end{equation}
We find that two largest eigenvalues
\begin{equation}
\lambda^{(1)}\approx 9.95, 
\quad \lambda^{(2)}\approx 3.82,
\label{lambda12}
\end{equation} 
are significant in comparison with the RMT prediction.
That is, they are significantly above the upper bound $\lambda_+$
and, therefore, reflect the non-random structure of the data. 

\bigskip\noindent
{\bf Identification of Two Dominant Factors}
\medskip

We have demonstrated that two largest eigenvalues are significant.
In Figure \ref{fig:e12new}, the components of the eigenvectors $\bm{V}^{(1,2)}$ 
are shown for production, shipments, and inventory by type of goods.
Taking advantage of this information revealed by disaggregated data, we can identify two
eigenvectors as follows.

We observe that the first eigenvector $\bm{V}^{(1)}$ captures the comovements of production and shipments. Thus, the first dominant factor corresponding to the largest eigenvalue is the ``aggregate demand'' which moves both shipments and production in all the sectors. This is nothing but Keynes' principle of effective demand.
\begin{quote}
The central Keynesian proposition is not nominal price rigidity but the principle of effective demand (Keynes, 1936, Ch.3). In the absence of instantaneous and complete market clearing, output and employment are frequently constrained by aggregate demand. In these excess-supply regimes, agents' demands are limited by their inability to sell as much as they would like at prevailing prices. Any failure of price adjustments to keep markets cleared opens the door for quantities to determine quantities, for example real national income to determine consumption demand, as described in Keynes' multiplier calculus. $\cdots\cdots$ In Keynesian business cycle theory, the shocks generating fluctuations are generally shifts in real aggregate demand for goods and services, notably in capital investment. \citep{tobin1993price}
\end{quote}
The textbook multiplier analysis depends, of course, on the assumption that changes in shipments induce almost equal changes in production.

It is noteworthy that in 15 out of 21 cases, responses of inventory stock is negative, suggesting depletion of inventory stock in response to the first dominant factor $\bm{V}^{(1)}$  which we interpret as the aggregate demand. When a firm's production is subject to increasing marginal cost and/or production takes time, there is a buffer-stock motive for inventory. The buffer stock motive produces the negative correlation between inventory investment and sales when demand exogenously changes \citep{hornstein1998iib}.

The second dominant factor $\bm{V}^{(2)}$, on the other hand, captures the comovements of production and inventory. Its effect on shipments is relatively small. 
\citet{metzler1941nature}'s classical work demonstrates that production is made not only for the purpose of meeting the current sales or shipments but also for maintaining the firm's desired level of inventory stocks. The firm's desired level of inventory stocks depends on {\it anticipations\/} of future demand or sales.
That is, inventory investment becomes, in part, ``speculative''.
In fact, based on a robust stylized fact that the variance of production is greater than that of shipments, 
\citet{blinder1991taking} 
and others have demonstrated that the role of inventory stock is not confined to production smoothing. Rather, they show that inventories apparently destabilize output at the macro level. Whether originated from the Metzlerian inventory accelerator (the idea that desired inventory stocks rise with sales) or from the $(S, s)$ inventory strategy at the micro level, inventory and production tend to move together. Our second dominant factor $\bm{V}^{(2)}$ appears to capture this channel. For convenience, we call it ``inventory adjustments''.

Whether direct or indirect, $\bm{V}^{(1)}$ and $\bm{V}^{(2)}$ are basically demand factors.
It is difficult to interpret $\bm{V}^{(1)}$ and $\bm{V}^{(2)}$ as ``productivity shocks'' for the following reason.
Changes in total factor productivity (TFP) emphasized by Real Business Cycle theory such as \citet{kydland1982time} bring changes in marginal cost in production. In order to minimize costs, the firm increases (decreases) production and accumulates (reduces) inventories during times when marginal cost is low (high). Thus, not only final demand (consumption / investment) usually analyzed by RBC but inventory investment also moves together with production when productivity changes \citep{hornstein1998iib}. In other words, productivity shocks, temporary shocks in particular, must move all the variables, namely production, shipments, and inventory in tandem. Figure \ref{fig:e12new} clearly shows that $\bm{V}^{(1)}$ captures the comovements of production and shipments while $\bm{V}^{(2)}$ the comovements of production and inventory, leaving the third variable relatively unaffected in both cases. The two dominant factors, therefore, do not represent productivity shocks. 

\section{Dominant Eigenmodes and the Business Cycles}\label{sec:eigen}

So far, we have found that 
(1) the $T=60$ (months) and the $T=40$ fluctuations are significant, and also that
(2) two largest eigenvalues, $\lambda^{(1,2)}$ of the correlation matrix,
are significant based on the RMT.

\def\ta{\widetilde{a}}
These two findings suggest that the contribution of two largest eigenvalue components at $T=60$ and $T=40$ periods are properly defined as the ``business cycles''. In this section, we measure the contribution of this ``business cycle'' factor in economic fluctuations. Toward this goal, we express the normalized growth rate in Fourier modes of the eigenvectors.

First, we decompose the normalized growth rate $w\ga(t_j)$ 
on the basis of the eigenvectors $\bm{V}^{(n)}$;
\begin{equation}
w\ga(t_j)=\sum_{n=1}^{M} a_n(t_j)\, V\ga^{(n)}.
\label{d1}
\end{equation}
By substituting this into Eq.~(\ref{Cdef}) and comparing it with Eq.~(\ref{CVV}),
we find that
\begin{equation}
\langle a_n a_{n'}\rangle_t = \delta_{nn'}\lambda^{(n)}.
\label{lamaa}
\end{equation}

The Fourier decomposition of the coefficients $a_n(t_j)$ is
\begin{equation}
a_n(t_j)=\frac1{\sqrt{\tn}}\sum_{k=0}^{\tn-1}\ta_n(\omega_k) \,e^{-i \omega_k t_j},
\label{fd2}
\end{equation}
similar to Eq.~(\ref{fd}). 
This leads to
\begin{equation}
\lambda^{(n)}=\frac{1}{\tn}\sum_{k=0}^{\tn-1}\lambda^{(n)}(\omega_k),
\quad
\lambda^{(n)}(\omega_k):=\left|\ta_n(\omega_k) \right|^2.
\label{lnln}
\end{equation}

From Eq.~(\ref{fd}) and Eq.~(\ref{lnln}), we obtain
\begin{equation}
\tw\ga(\omega_k)=\sum_{n=1}^{M}\ta_n(\omega_k)\, V\ga^{(n)}.
\end{equation}
Substituting this into Eq.~(\ref{defps}),
we obtain that the following decomposition of the power spectrum
\begin{equation}
p(\omega_k)=\frac{1}{M}\sum_{n=1}^{M}\lambda^{(n)}(\omega_k).
\end{equation}
By summing the both-hand sides over $k$, we recover Eq.~(\ref{tracec}):
\begin{equation}
1=\frac{1}{N'}\sum_{k=1}^{N'-1}p(\omega_k)=\frac{1}{M}\sum_{n=1}^{M}\lambda^{(n)}.
\end{equation}
This shows that the eigenvectors corresponding to the larger eigenvalues have larger contribution
to the power spectrum {\it on average}: For example, the relative contribution 
of the first eigenvector is 16\% as $\lambda^{(1)}/M\approx 0.16$ ($M=63$).
If the contributions of all the eigenvectors were the same, 
it would be only 1.6\% ($1/M\approx 0.016$).
The sum of the relative contribution of the first and the second eigenvector
components amounts to 21\%.
Dominance of two large eigenvalues is clear.

So far, we have discussed the average contribution.
The important question is in what part of the power spectrum the large eigenvalues
have significant contributions.
Figures~\ref{fig:powerspectrum_eigenmodes} and \ref{fig:powerspectrum_eigenmodes_ratio}
answer this question.
In Figure~\ref{fig:powerspectrum_eigenmodes}, the dark gray spectrum at the bottom is the
contribution of the first eigenvector, $\lambda^{(1)}(\omega_k)/M$, 
whereas the spectrum with lighter shading is that of the second eigenvector, $\lambda^{(2)}(\omega_k)/M$.
The solid black line is the whole spectrum, $p(\omega_k)$, which is the same as
the spectrum with gray shading in Figure~\ref{fig:powerspectrum}.
We observe that the eigenmodes of the larger eigenvalues
tends to have greater contributions as $T$ gets larger.

This is better illustrated in
Figure~\ref{fig:powerspectrum_eigenmodes_ratio} which shows
the relative contribution in each bin of period with the
same notations as used in Figure~\ref{fig:powerspectrum_eigenmodes}.
We observe here that the relative contribution of the first eigenmode 
is about 20\% for $T\gtrsim 40$ whereas it is only 10\% or less for $T\lesssim 40$.
The same is true for the second eigenmode.
The sum of the relative contribution of the first and second eigenmodes is 
about 40\% for $T\gtrsim 38$ but is only 20\% for $T\lesssim 38$. 
(The boundary $T\approx38$ is denoted by the vertical dashed line and the mark 
``A" on the horizontal axis.)
This is another reason why we rejected smaller periods like $T=24$
as insignificant in Section 3;
The relative contributions of the dominant two
eigenmodes for $T\lesssim 38$ are much smaller than those for $T=40$ and $T=60$.
It suggests that fluctuations in shorter periods $T<38$ are basically due to noises,
and do not reflect of any systematic comovements.

We have identified the two dominant factors $\bm{V}^{(1)}$ and $\bm{V}^{(2)}$ as demand factors, not productivity shocks. It is important to recognize that the fact that the relative contribution of $\bm{V}^{(1)}$ and $\bm{V}^{(2)}$ is roughly 40\% does not mean that the residual 60\% is to be explained by other systematic factors such as ``productivity shocks'' emphasized by RBC. The residual 60\% is to be explained by $\bm{V}^{(j)}$ ($j=3,4,\ldots$), but the significance of $\bm{V}^{(j)}$ ($j=3,4,\ldots$) is rejected by the RMT (Section~\ref{sec:rmt}). The residual 60\% is, therefore, not really a part of systematic ``business cycles'', but rather caused by various unidentified random factors. In fact, \citet{slutzky1937src} suggested that the summation of pure random shocks might generate apparently ``cyclical'' fluctuations; In effect, he rejected the existence of systematic ``business cycles''. Contrary to his assertion, our findings confirm the presence of business ``cycles'' which {\it systematically\/} account for 40\% of economic fluctuations. Such ``business cycles'' defined based on $\bm{V}^{(1)}$ and $\bm{V}^{(2)}$ at $T=40$ and $60$ are basically caused by demand factors.
In the next section, we will investigate the cyclical behavior of production, shipments and inventory in greater detail.

\section{Production and Shipments}\label{sec:ps}
In the previous sections, we have shown the presence of 
the two eigenmodes of the correlation matrix for the $T=40$ and $60$ Fourier modes.
Let us now recover the behavior of production and shipments in these dominant modes.
In order to do so, we first express the decomposition of the IIP data
to each Fourier and eigenmode by combining Eqs.~(\ref{d1}) and (\ref{fd2}):
\begin{align}
w\ga(t_j)&=\frac1{\sqrt{\tn}} \sum_{n=1}^{M} \sum_{k=0}^{\tn-1}
\ta_n(\omega_k) \,e^{-i \omega_k t_j} V\ga^{(n)}
\label{domi0}\\
&=\frac2{\sqrt{\tn}} \sum_{n=1}^{M} \sum_{k=1}^{(\tn-1)/2}
\Re\left(\ta_n(\omega_k)e^{-i \omega_k t_j}\right) V\ga^{(n)}.
\label{domi}
\end{align}
Here we used the facts that $\tilde{a}_n(0)=0$ due to $\tw\ga(0)=0$,
$\tilde{a}_n(\omega_k)=\tilde{a}_n^*(\omega_{N'-k})$ due to $a_n(t_j)$ being real,
and that $N'-1=238$ is even.
The ``business cycle component'' is obtained by limiting the above sum to over only
$k=4$ ($T=60$) and $k=6$ ($T=40$) and $n=1, 2$.

Figure~\ref{fig:T6040_2modes} shows the cyclical movements
of production, shipments, and inventory. As explained in the previous section,
they are defined as cycles produced by two dominant factors at $T=40$
and $T=60$. The cycles shown in Figure~\ref{fig:T6040_2modes} are the
average values over the goods, $w_\alpha (t_j):=\sum_{g=1}^{21}w\ga(t_j)/21$
for $T=60$ (the left panel) and for $T=40$ (the right panel).
The top row in each panel is for the first dominant factor,
the middle row for the second dominant factor,
and the bottom row for their sum.

In the first eigenmode (the top row), production (the solid line) and
shipments (the dashed line) move together with equal amplitude.
In contrast, in the second eigenmode (middle)
the movements of shipments are almost flat.
This is consistent with the results in Figure~\ref{fig:e12new}.

Recall that we identify the first eigenmode with the aggregate demand 
while the second eigenmode with inventory adjustments.
It is notable that although in each eigenmode, production and shipments
move with the same (or opposite) phase as is guaranteed by our expansion,
they are shifted in time once the sum over two dominant eigenmodes is done.
This is because of the difference in their amplitudes.

For the $T=60$ oscillations, production is delayed by the second 
eigenmode as shown by two black arrows in Figure~\ref{fig:T6040_2modes}.
In contrast, shipments receive almost no contribution from the second eigenvector. 
As a result, production is delayed a little after shipments.
We can exactly derive this delay-time.
From decomposition (\ref{domi0}), we see that
the average of the normalized growth rates
$w_{\alpha, g}$ over the goods ($g=1,\dots21$),
$\bar{w}_\alpha(t_j):=\sum_{g=1}^{21}w\ga(t_j)/21$,
contains the following $T=60$ ($k=4$) part:
\begin{align}
\bar{w}_\alpha(t_j)=
\frac1{\sqrt{\tn}} 
\left[\ta_1(\omega_4) \bar{V}_\alpha^{(1)}+\ta_2(\omega_4) \bar{V}_\alpha^{(2)}\right] 
e^{-i \omega_4 t_j} +\cdots .
\end{align}
where $\bar{V}_\alpha^{(n)}:= \sum_{g=1}^{21}V_{\alpha,g}^{(n)}/21$.
Therefore, denoting the absolute value and the phase of the coefficients as,
\begin{equation}
\ta_1(\omega_4) \bar{V}_\alpha^{(1)}+\ta_2(\omega_4) \bar{V}_\alpha^{(2)}
:=A_\alpha e^{i\phi_\alpha},
\end{equation}
where $A_\alpha>0$,
we find that production ($\alpha=1$) and shipment ($\alpha=2$)
obey the following relations:
\begin{align}
\bar{w}_1(t_j)&=A_1 e^{-i(\omega_4 t_j-\phi_1)}+\cdots,\\
\bar{w}_2(t_j)&=A_2 e^{-i(\omega_4 t_j-\phi_2)}+\cdots.
\end{align}
This means that the delay-time of production behind shipments for the period $T=60$,
$\Delta^{\rm (SP)}_{60}$, is given by
$(\phi_1-\phi_2)/\omega_4$.
Because $\omega_j=2\pi j/N'$ [1/month] with $N'=239$, we obtain
the delay-time $\Delta^{\rm (SP)}_{60}$ in units of months as follows:
\begin{equation}
\Delta^{\rm (SP)}_{60}:=
\frac{239}{4}\frac{1}{2\pi}
\text{Arg}\left[
\frac{\ta_1(\omega_4)\bar{V}_1^{(1)}+\ta_2(\omega_4)\bar{V}_1^{(2)}}
{\ta_1(\omega_4)\bar{V}_2^{(1)}+\ta_2(\omega_4)\bar{V}_2^{(2)}}
\right]\approx 4.28.
\label{dsp60}
\end{equation}

It is clear in this expression that the lead-lag between shipments and production
is caused by the combination of two eigenmodes.
For inventory, the shift occurs in the opposite direction,
and as a result, it lags by 
\begin{equation}
\Delta^{\rm (PI)}_{60}\approx 10.95,
\label{dpi60}
\end{equation}
behind production.

The basically same story holds for the $T=40$ oscillations (Figure~\ref{fig:T6040_2modes}).
We thus obtain$^7$
\begin{equation}
\Delta^{\rm (SP)}_{40}\approx 1.93, \quad \Delta^{\rm (PI)}_{40}\approx 13.62.
\label{dsp40}
\end{equation}

We can formally test whether the time-delays $\Delta^{\rm (SP)}_{60}$ and $\Delta^{\rm (SP)}_{40}$
are statistically significant or not.
Toward this goal, we randomize the rest of the eigenmodes and evaluate those values.
The simulation is done in following steps:
\begin{enumerate}
\item Subtract the eigenmodes of the first two eigenvalues from the actual data to define:
\begin{equation}
w\ga'(t_j):=
w\ga(t_j)-\sum_{n=1}^{2} a_n(t_j)\, V\ga^{(n)}.
\end{equation}
(see Eq.~(\ref{d1})).
\item Randomly reorder the time-series $w\ga'(t_j)$ separately,
{\it i.e.}, with randomization different for each $\alpha$ and $g$.
\item Combine the reshuffled part with the first two eigenmodes to create a simulated data.
\item Calculate the eigenvalue distribution and see whether the largest two
eigenvalues still remain. If they are, we proceed to the next step.
(Note that the above randomization process changes the eigenvalues.)
\item Calculate $\Delta^{\rm (SP)}_{60}$ and $\Delta^{\rm (SP)}_{40}$
for the resulting simulated data.
\end{enumerate}
In this manner, the noise part (the eigenmodes except for the first and second eigenmodes) 
are reshuffled. We then check whether the values we
obtain for the time-delays, $\Delta^{\rm (SP)}_{60,40}$ and $\Delta^{\rm (PI)}_{60,40}$,
stand out against the noises. 
We carried out this simulation $10^5$ times.
The result obtained is that the first 
two eigenvalues always exist as seen in Figure~\ref{fig:simulation}.
The distribution of the time-delays are shown in Figure~\ref{fig:simulation2}
with the summary statistics of simulation results in Table \ref{tab:suchi}.
We observe that the delay-time obtained
in Eqs.~(\ref{dsp60}) and (\ref{dsp40})
are statistically significant.

In Appendix~B, we perform cross-spectrum analysis and estimate the
phase in the spectrum to obtain the lead-lag relationships among
production, shipments and inventory. For the series of the sum of the
first and second eigenmodes, the results obtained by the
cross-spectrum analysis are consistent with, and thereby confirm from
a different angle, those presented in this section. In addition, by
comparing them with the cross-spectrum for the original data, we can say that
the extraction of the two dominant eigenmodes or our definition of
``business cycles'' is essential for understanding the lead-lag
relationship between shipments and production.
 
The sum of the components of the first
and the second eigenvectors at the $T=40$ and $T=60$ periods is
compared with the original data for production, shipments and
inventory in Figure~\ref{fig:mvreal}. To make comparison easier, we
applied a simple moving average operation to the original data over
one-year period. For the original data so smoothed out, we can hardly
detect any lead-lag relation between production and shipments.  In
contrast, in the extracted ``business cycles'', shipments clearly
precede production with the average lead time of four months.

In Figure~\ref{fig:mvreal}, we observe a close one-to-one
correspondence between the original data and the extracted business
cycles. The peaks and troughs of two cycles coincide very well
although their amplitudes slightly differ due possibly to economic
fluctuations with longer cycles ($T>60$) and external shocks. A linear
combination of two sinusoidal functions oscillating at different
periods gives rise to an extended cyclic motion with the period of
$T=120$. Figure~\ref{fig:mvreal} thus spans two cycles of the
endogenous business fluctuations. The characteristic behavior of the
original data shown in the bottom panel is well captured by the
``business cycles'' defined here.

Japan had experienced a long stagnation beginning 1990 through
2002. One might consider a possible ``structural break'' during the
``lost decade". However, the finding of such steady cycles shown in
Figure~\ref{fig:mvreal} suggests that the Japanese economy has had actually a fairly
stable structure over the last two decades. Besides, our sample period
(January, 1988--December, 2007) basically covers the whole period of the long stagnation.
Thus, rather than chopping our data into sub-sample periods, we resort to
the out-of-sample examination of the 2008-09 economic crisis in the next section.

We have found that production lags behind shipments
over the extracted ``business cycles". It does so inventory adjustments tend to
lag behind the initial changes in shipments. Suppose that shipments or
final demand increased in the economy as a whole. It induces a
contemporaneous increase of production, and at the same time,
depletion of inventory stock in many sectors and also for many types
of goods; This is well captured by the first eigenmode which we
interpret the ``aggregate demand'' (the top panel of
Figure~\ref{fig:T6040_2modes}). Now, an increase of final demand and
depletion of inventory stock make firms revise their expectations on
future demand, and accordingly the level of desired inventory stock.
If demand does not increase further, firms produce more for the
purpose of inventory accumulation. This is captured by the second
eigenmode which we interpret the ``inventory adjustments'' (the middle
row of Figure~\ref{fig:T6040_2modes}). As \citet{metzler1941nature}
shows, inventory adjustments take time. This makes production lag
behind shipments. Our analysis shows that the average lag is about 4
months. In the case of inventory stock, an {\it increase\/} of
shipments initially causes {\it depletion\/} of inventory, and,
therefore, the lag becomes even longer than in the case of production.
We find that the average lag of inventory behind shipments is about 11
months. In this way, we can easily understand that shipments lead
production and inventory within the framework in which changes in
real demand or shipments are the major shocks.
Note that productivity shocks emphasized by RBC, by
definition, instantaneously change production. There is no room for
production to lag behind shipments in such framework.

\section{The 2008-09 Economic Crisis}

The 2008-09 financial crisis triggered by the bursting of the housing bubbles in the U.S.\ immediately spread all over the world, and caused great damage on the global economy, reminiscent of the Great Depression. 
Based on the analyses in the previous sections, 
we update the IIP data to cover the period from January 2008 ($j=241$) through June 2009 ($j=258$), and perform an out-of-sample test.

The extended data shown in Figure~\ref{fig:crisis1new} demonstrates a dramatic drop of production in Japan. Panel (a) of Figure~\ref{fig:crisis_abs} shows the volatility $P(t)$ of the IIP defined by 
\begin{equation}
P(t):=\sum_{\alpha=1}^{3}\sum_{g=1}^{21}|w_{\alpha,g}(t)|^2.
\label{IIP_volatility}
\end{equation}
We observe the exceptionally large amplification of $P(t)$ during the 2008-09
economic crisis.

This situation, therefore, provides us with a perfect opportunity for out-of-sample test for validity of the dominant eigenmodes that we have identified using the random matrix theory. 
To this end, we decompose the extended IIP data based on the exactly same eigenvectors as in Eq.~(\ref{d1}). Then $P(t)$ is expressed in terms of the mode amplitude coefficients $a_{n}(t)$ as follows:
\begin{equation}
P(t)=\sum_{n=1}^{63}|a_{n}(t)|^2.
\label{IIP_volatility1}
\end{equation}
Panels (b) and (c) of Figure~\ref{fig:crisis_abs} show the partial contributions of the largest and the second largest modes to $P(t)$, respectively. Plainly, an unprecedented rise of 
volatility of the first dominant factor, namely $|a_1(t)|^2$ accounts largely for a rise of $P(t)$ during the 2008-09 crisis.
The contribution of the second dominant factor is not so great as that of the first dominant factor.

The relative contributions of the first and second eigenmodes defined by
\begin{equation}
\pi_{n}(t):=\frac{|a_{n}(t)|^2}{P(t)}.
\label{IIP_egmode_rs}
\end{equation}
are shown in Figure~\ref{fig:crisis_rel}. 
The total sum of $\pi_{n}(t)$ with respect to $n$ is normalized to unity. 
The first dominant eigenmode accounts for almost a half of the total power of fluctuations during the 2008-09 recession. Note that the {\it relative\/} contribution of the first eigenmode, namely about one half, is comparable to that for the in-sample period.
This result demonstrates that our factor model holds up over the 2008-09 crisis despite of the fact that a fall of industrial production during the period as shown in Figure \ref{fig:crisis1new} was truly unprecedented.

We have seen that the first eigenmode basically explains a fall of industrial production during 2008-09. Now, Figure \ref{fig:crisis11} shows IIP and the index of exports during the period under investigation. Plainly, production fell in response to a sudden decline in exports. We emphasize that a fall of exports was caused by a sudden shrinkage of the world trade due to the financial crisis accompanied by the global recession, and that {\it it was basically exogenous to the Japanese economy\/}. In Section \ref{sec:rmt}, by analyzing the components of eigenvectors (Figure \ref{fig:e12new}), we identified the first dominant factor as the aggregate demand. To the extent that we identify a sudden fall of exports, an exogenous negative aggregate demand shock, as the basic cause of the 2008-09 recession, the dominance of the first eigenmode provides us with another justification for our interpreting it as the aggregate demand factor. This confirms once again the validity of the Old Keynesian View \citep{tobin1993price}.

\section{Concluding Remarks}\label{sec:concl}

In his comments on \citet{shapiro1988sbc} entitled ``Sources of Business Cycle Fluctuations,'' \citet{hall1988csw} made the following remark: 
\begin{quote}
All told, I find the results of this paper harmonious with the emerging middle ground of macroeconomic thinking. There are major unobserved determinants of output and employment operating at business-cycle and lower frequencies. Some are technological, some are financial, some are monetary \citep[p.151]{hall1988csw}.
\end{quote}

For finding out such unobserved determinants of business cycles, the standard identifying restriction is that only technology shocks affect output permanently. As discussed in Introduction, we maintain that this assumption is untenable. Using production, shipments, and inventory data, we analyze the causes of business cycles more straightforwardly than in the standard approach.

We have identified (i) two periods $T=60$ and $40$ months for the Japanese business cycles, and also (ii) the two significant factors causing comovements of IIP across 21 goods. Taking advantage of the information revealed by disaggregated data, we can reasonably interpret the first factor as the aggregate demand, and the second factor as intentional inventory adjustments. It is difficult to interpret the two dominant factors as productivity shocks. Within the framework of the standard RBC, productivity shocks are expected to affect production, shipments, and inventory together. The two dominant factors we have identified do not fulfill this condition (Figure \ref{fig:e12new}). 

We have also found that, for the part of shipments and production explained by two factors, shipments lead production by a few months. This lead-lag relationship between shipments and production caused by inventory adjustment is not clearly seen in the original data. However, inventory adjustments which generate this lead-lag relationship between shipments and production are an important part of ``business cycles'' defined by two dominant factors.

Furthermore, out-of-sample analysis of the 2008--09 recession demonstrates that the first dominant factor largely explains this major episode. We know that it was basically caused by a sudden fall of exports beginning the fourth quarter of the year 2008 (Figure \ref{fig:crisis11}). Therefore, to the extent that exports can be reasonably regarded as exogenous {\it real} aggregate demand shocks to the Japanese economy, we can identify the first dominant factor as the real aggregate demand. All these findings suggest that the major cause of business cycles is {\it real demand\/} shocks accompanied by significant inventory adjustments.

It is important to recognize that this conclusion of ours does not necessarily mean that technology shocks are unimportant. On the contrary, it is very plausible that a major driving force of fixed investment is technology shocks.
Furthermore, one might argue that technology shocks is also a driving force of consumption by way of introduction of new products \citep{aoki2002dsc}.
The point is that technology shocks affect the macroeconomy by way of changes in aggregate demand, notably investment, in the short-run. For the purpose of studying business cycles, it is wrong to consider technology shocks as direct shifts of production function.

The conclusion we drew from our analyses of the Japanese industrial production data is clear. To study whether the same results hold up for other economies is obviously an important research agenda. We believe that the methods we employed in the present paper can be usefully applied to the study of business cycles in other economies.

Finally, we have identified 40 to 60 months ``cycles'' of aggregate demand. The theory of such ``cycles'' of aggregate demand is beyond the scope of the present paper. After all, the multiplier-accelerator model of \citet{Samuelson1939} and \citet{hicks1951ctt} may be an answer. The problem awaits further investigation.

\section*{Acknowledgments}

We gratefully acknowledge detailed comments made by two anonymous referees and the editor's kind suggestions.
This work was supported in part by {\it the Program for Promoting Methodological Innovation in Humanities and Social Sciences by Cross-Disciplinary Fusing\/} of the Japan Society for the Promotion of Science.

\newpage

\appendix
\renewcommand{\thesection}{Appendix \Alph{section}}
\renewcommand{\theequation}{\Alph{section}.\arabic{equation}}

\section{Significance of Dominant Factors in Finite Sample}\label{sec:appendb}
\setcounter{equation}{0}

We have found that two large eigenvalues exist outside the range of the RMT prediction
for the covariance matrix $\bm{C}$ of IIP.
This result, however, holds exactly true only asymptotically in the limit (\ref{limits}),
while we have only finite sample, $N'=239$ and $M=63$.
Also the RMT applies theoretically to iid data as noted in Section \ref{sec:rmt}.
In this appendix, we will first show that nontrivial 
autocorrelation exists for the growth rate, and then present a simulation result
which shows that in spite of the ``finite-size" effect and the nontrivial autocorrelation,
the dominant factors we have found are still significant.

The autocorrelation $R(t)$ of the normalized 
growth rate $w_{\alpha,g}$ is defined by the following:
\begin{equation}
R_{\alpha, g}(t_m):=\frac{1}{N'-m}\sum_{j=1}^{N'-m} w_{\alpha,g}(t_j) w_{\alpha,g}(t_{j+m}).
\label{auto1}
\end{equation}
By definition, $R_{\alpha, g}(0)=1$, and 
if there are no autocorrelations, $R_{\alpha, g}(t_m)=0$ for $m \ge 1$.
Figure \ref{fig:autocorrelation} shows the autocorrelation averaged over the goods;
\begin{equation}
\bar{R}_{\alpha}(t_m):=\frac{1}{21}\sum_{g=1}^{21} R_{\alpha, g}(t_m).
\label{auto2}
\end{equation}
We observe that for both production ($\alpha=1$) and shipments ($\alpha=2$),
the autocorrelations have nontrivial values for $t=1$ month: $R_1(1)\approx-0.31$
and $R_2(1)\approx-0.39$, while we have $R_3(1)\approx0.007$ for inventory.

Since this nontrivial autocorrelation,
in addition to the finiteness of the data size, 
may bring corrections to the RMT results,
we have done a simulation in the following steps:
\begin{enumerate}
\item[(1)]
Random ``rotation" of each time-series,
\begin{equation}
w_{\alpha,g}(t_j)\rightarrow w_{\alpha,g}(t_{{\rm Mod}(j-\tau,N')})
\end{equation}
where $\tau\in[0,N'-1]$ is a (pseudo-)random integer, different for each $\alpha$ and $g$.
\item[(2)]
Then we have calculated the eigenvalues $\lambda^{(n)}$ for the resulting
set of the randomized data.
\end{enumerate}
The data randomized in the above manner has the same autocorrelation
and the same finite-size with the original data, but
correlations between production, shipments, and inventory of different goods
are destroyed. 

We have run this calculation $10^5$ times
and have obtained PDFs shown in Figure~\ref{fig:ringsim}.
In the plot (a) of Figure~\ref{fig:ringsim}, the gray area shows the RMT prediction
identical to that of Figure~\ref{fig:eigenvalue_distribution} and
the dots connected by solid lines show the result of the simulation.
We observe here that the simulation result differs from that of the RMT
prediction, but not to the extent that the values we have found in
actual IIP data, $\lambda^{(1)}\approx9.95$ and $\lambda^{(2)}\approx 3.82$
are nontrivial.
The same can be said from the plot (b), where the PDFs of the
first (largest) eigenvalue $\lambda^{(1)}$ and the second (next-largest)
eigenvalue $\lambda^{(2)}$ are given. The average of
$\lambda^{(1)}$ obtained from this simulation is $\bar{\lambda}^{(1)}\approx 2.47$
with standard deviation $\sigma_{\lambda^{(1)}}\approx 0.10$ and
$\bar{\lambda}^{(2)}\approx 2.29$, $\sigma_{\lambda^{(2)}}\approx 0.07$.

Therefore, we conclude that
the two largest eigenvalues $\lambda^{(1)}$ and $\lambda^{(2)}$ 
we have found in the actual IIP data are statistically significant factors,
which represent comovements of production, shipments, and inventory of different goods.

\section{Lead-Lag Relationships by Cross-Spectrum}\label{sec:xspec}
\setcounter{equation}{0}

In Section~\ref{sec:ps}, we evaluate the lead-lag relationships among
production, shipments and inventories by defining $\Delta^{\rm (SP)}$
and $\Delta^{\rm (PI)}$ for two cycles $T=60$ and 40.
In this appendix, we resort to the method of cross-spectrum
(see e.g. \citet{jenkins1968saa,bloomfield2000fat})
to identify the lead-lag relationships in terms of
coherency and phase.

For a pair of time-series of length $\tn$, $x(t_j)$ and $y(t_j)$,
we consider the Fourier transforms $\tilde{x}(\omega_k)$ and
$\tilde{y}(\omega_k)$ defined by
$\tilde{x}(\omega_k)=\sum_{j}x(t_j)e^{i\omega_k t_j}/\sqrt{\tn}$
as Eq.(\ref{fd}).
Their periodograms are given by
$I_{xx}(\omega_k):=|\tilde{x}(\omega_k)|^2$ and
$I_{yy}(\omega_k):=|\tilde{y}(\omega_k)|^2$, respectively. The
cross-periodogram for the pair is defined by
$I_{xy}(\omega_k):=\tilde{x}(\omega_k)^*\,\tilde{y}(\omega_k)$ which
is a complex number (except $\omega=0,\pi$).

Just as the periodogram is smoothed to obtain an estimate of spectrum,
the cross-periodogram is smoothed to obtain estimate of
cross-spectrum, i.e.
\begin{equation}
  \hat{s}_{xy}(\omega_k)=\sum_{\ell}\gamma_{\ell}\,I_{xy}(\omega_k-\omega_{\ell})
\label{xspec_smooth}
\end{equation}
where $\gamma_{\ell}$'s are the weights for smoothing.
Similarly, we can obtain the auto-spectra,
$\hat{s}_{xx}$ and $\hat{s}_{yy}$. Then, the ratio defined as
\begin{equation}
  \hat{\kappa}_{xy}^2(\omega_k)=
  \frac{|\hat{s}_{xy}(\omega_k)|^2}{\hat{s}_{xx}(\omega_k)\,
    \hat{s}_{yy}(\omega_k)}
\end{equation}
is the estimate of squared coherency. It satisfies the following inequality:
\begin{equation}
  0\leq\hat{\kappa}_{xy}^2(\omega_k)\leq 1,
\end{equation}
with 0 corresponding to no linear
dependence and 1 corresponding to exact linear dependence of the
two time-series at frequency $\omega_k$.

The cross-spectrum can be expressed
in terms of its amplitude and phase as follows:
\begin{equation}
  \hat{s}_{xy}(\omega_k)=|\hat{s}_{xy}(\omega_k)|\,
  e^{2\pi i\hat{\phi}_{xy}(\omega_k)}.
\end{equation}
The phase $\hat{\phi}_{xy}(\omega_k)$ estimates the lead-lag
relationship between the series. Note that the estimate of the phase
becomes meaningless when the cross-spectrum is small in its
amplitude.

In the following estimation, we consider the multiple time-series
consisting of the first two components, namely
$w\ga(t_j)=\sum_{n=1}^{2} a_n(t_j)\, V\ga^{(n)}$, averaged over the
index $g$ of goods for $\alpha=1,2,3$. We take two pairs of
shipments and production ($\alpha=(2,1)$), and production and inventory
($\alpha=(1,3)$) for $x(t_j)$ and $y(t_j)$. For the smoothing of the
cross-periodograms, we used a modified Daniell kernel for the weight
$\gamma_{\ell}$ of length 11 (corresponding to the bandwidth
$\Delta\omega /2\pi=0.01226$ cycles per month) with no tapering. The standard formulae
of significance levels and confidence intervals are used in the
calculation \citep{bloomfield2000fat}.

Figure~\ref{fig:xspec} shows the estimates of
$\hat{\kappa}_{xy}^2(\omega)$ and $\hat{\phi}_{xy}(\omega)$ for the
pair (SP) of shipments and production (panel (a)) and for the pair
(PI) of production and inventory (panel (b)). For
$\hat{\kappa}_{xy}^2(\omega)$, the confidence levels (90\% and 99\%)
are depicted by dotted horizontal lines, which are calculated under
the null hypothesis of zero coherency at each frequency. The
estimate of $\hat{\kappa}_{xy}^2$ and its 95\% confidence interval
are shown by a curve and the shaded region around it. One can then see
that $\hat{\kappa}_{xy}^2$ is significant at all frequencies
for the pair (SP), while it is significant only at low frequencies
corresponding to $T\gtrsim 20$ months for the pair (PI).

In the frequency regions where the squared coherency
$\hat{\kappa}_{xy}^2(\omega)$ is significant, one can evaluate the phase and its
confidence interval in each plot of $\hat{\phi}_{xy}(\omega)$. We
examine the sign and magnitude of the phase to evaluate the
lead-lag relationship and its magnitude. For the pair of (SP), at the
two frequencies $\omega_{k=6},\omega_{k=4}$ corresponding to $T=40,60$
months, respectively, the phase is significantly positive meaning that
shipments lead production at these frequencies. Its magnitude can be
evaluated in months by calculating $\hat{\phi}_{xy}(\omega)/\omega$
and its confidence intervals. The results are
\begin{equation}
  \Delta^{\rm (SP)}_{40}=2.26\ [1.43,3.09],\quad
  \Delta^{\rm (SP)}_{60}=2.74\ [0.973,4.49],
\end{equation}
where the square brackets represent 95\% confidence intervals.
For the sake of comparison, in Figure~\ref{fig:xspec}, the results
presented in Section~\ref{sec:ps} are depicted as two points with
error bars at $T=40,60$ in the plot of $\hat{\phi}_{xy}(\omega)$.

For the pair of (PI), the technique of {\it alignment}
\citep{jenkins1968saa} was used to weaken the biasing effect due to
rapidly changing phase in the presence of large lead-lag. The two
series of production and inventory are aligned by translating one
series a distance of 8 months so that the peak in the
cross-correlation function of the aligned series is close to zero
lag. Then the results are
\begin{equation}
  \Delta^{\rm (PI)}_{40}=9.04\ [7.86,10.2],\quad
  \Delta^{\rm (PI)}_{60}=9.32\ [7.56,11.1].
\end{equation}

Just at in panel (a) of Figure~\ref{fig:xspec}, in panel (b)
for the pair of (PI), the results presented in
Section~\ref{sec:ps} are depicted as two points with error bars in the
plot of $\hat{\phi}_{xy}(\omega)$. The confidence intervals do not
overlap with the error bars, presumably due to the alignment problem,
but we found no large differences from the results.

It is interesting to compare the above results {\it based on two
 extracted eigenmodes, namely our ``business cycles''\/}, with the
cross-spectrum for the original data, that is, by retaining all the
eigenmodes. Figure~\ref{fig:xspec0} shows the estimates of the
cross-spectrum for the original data: (a) for the pair of (SP)
and (b) for that of (PI). One can make the following observations for
$T=40,60$.
\begin{itemize}
\item The squared coherency $\hat{\kappa}_{xy}^2(\omega)$ is
  significant for both pairs. However, the phase
  $\hat{\phi}_{xy}(\omega)$ is significant for the pair (PI) with a
  similar magnitude of phase, whereas for the pair (SP), it is not
  significantly different from zero at those periods $T$.
\end{itemize}
This observation demonstrates that our extraction of two dominant
eigenmodes by the method of RMT is essential to reveal the lead-lag
relationship between shipments and production which is not clearly
present in the original data. We also note that the extraction does
not deform the other results of significant coherency and the lead-lag
relationship between production and inventory.

We conclude that the method of cross-spectrum and that used in
Section~\ref{sec:ps} are consistent in the evaluation of lead-lag
relationships among production, shipments and inventory, and also that
the extraction of two dominant eigenmodes or our definition of
``business cycles'' is essential for understanding the lead-lag
relationship between shipments and production.

\newpage

\renewcommand{\thefigure}{\arabic{figure}}

\def\figp#1#2#3#4{
\begin{figure}[!tp]
\begin{center}
\includegraphics[width=#3\textwidth,bb=#4]{./figures/#1.pdf}
\caption{#2}
\label{fig:#1}
\end{center}
\end{figure}}

\begin{figure}[!tp]
\begin{center}
\includegraphics[width=0.8\textwidth,bb=208 27 565 454]{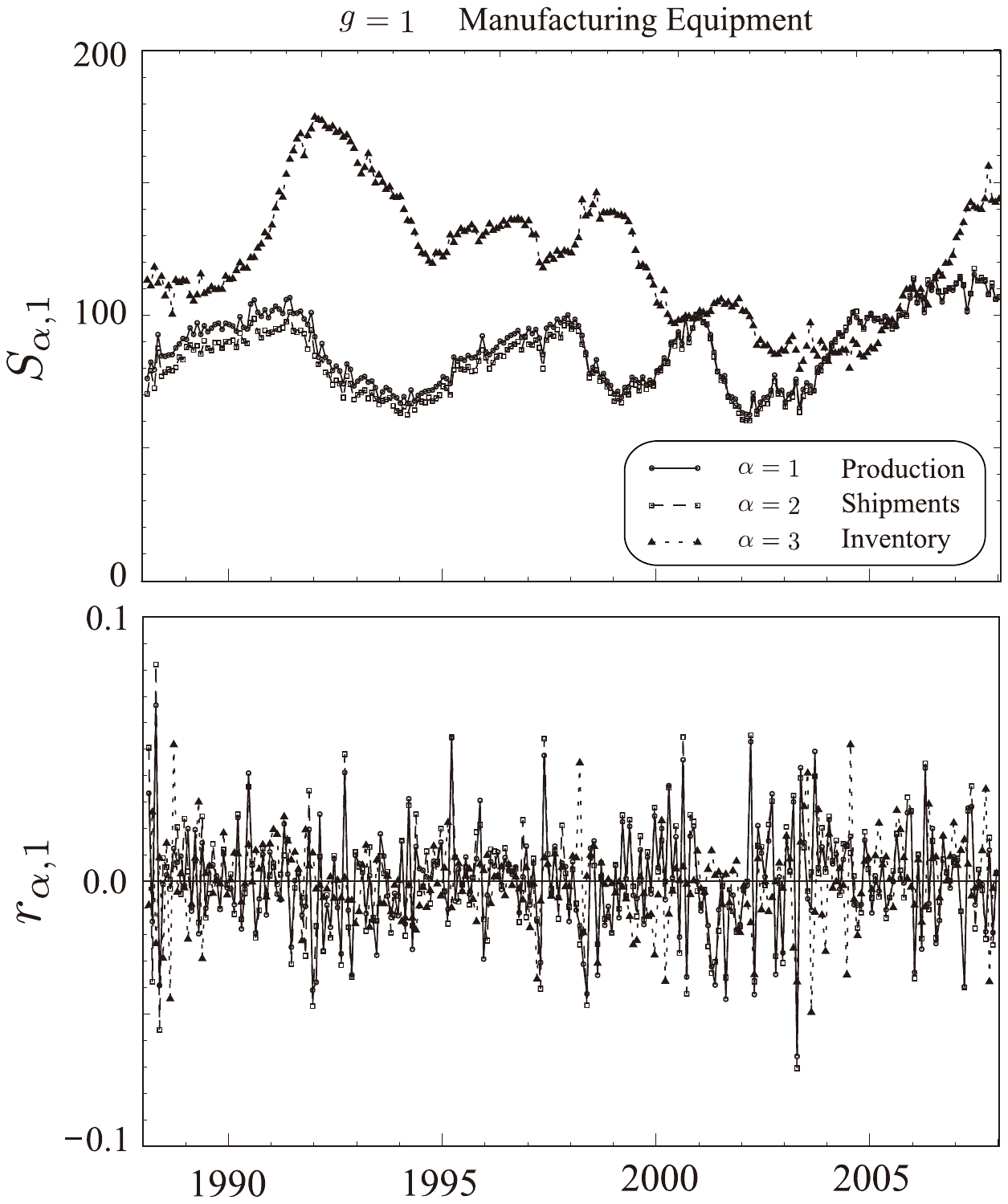}
\caption{A sample of the IIP data for $g=1$ (Manufacturing Equipment).
\protect\newline
Note: 
The upper panel shows the indices $S_{\alpha,1}(t)$ while
the lower panel shows the corresponding growth-rates $r_{\alpha,1}(t_j) $ defined by
Eq.(\ref{rdef1}).
As explained in the legend, 
$\alpha=1$ (Production)  is shown by open circles connected by solid lines,
$\alpha=2$ (Shipment)  is shown by open squares  connected by dashed lines,
and $\alpha=3$ (Inventory)  is shown by rectangles connected by dotted lines.
The horizontal axis shows the months beginning January 1988 through December 2007.}
\label{fig:data_g1new}
\end{center}
\end{figure}

\clearpage
\figp{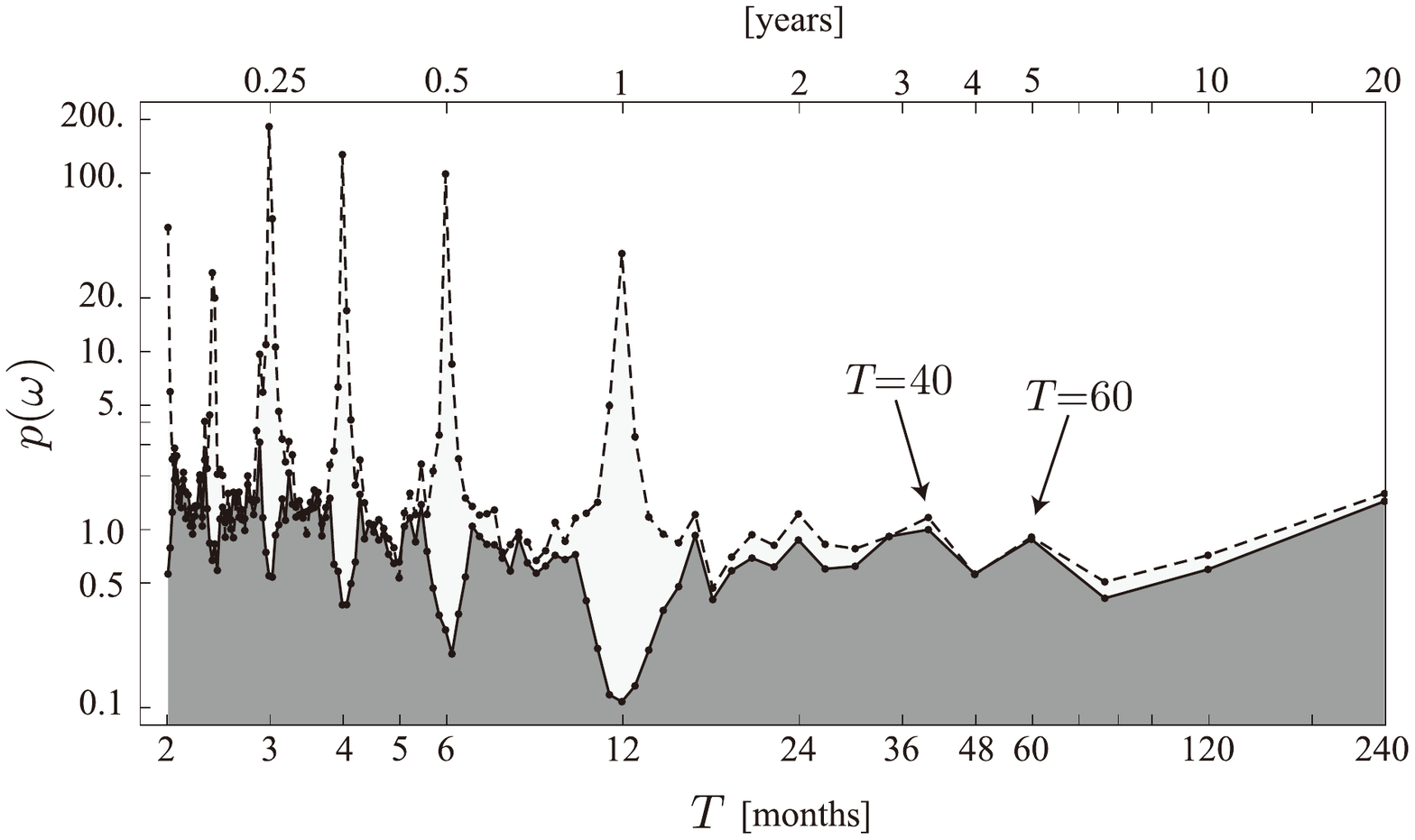}{The average power spectrum $p(\omega)$ of the normalized growth rate.
\protect\newline
Note: The horizontal axes is chosen to be the period, $T:=2\pi/\omega$,
which is shown in units of months at the bottom and in units of years at the top.
The points connected by solid lines (with darker shading) are the spectrum 
for the seasonally-adjusted data, while the points connected with dashed lines 
(with lighter shading) are for the non-adjusted original data.}{1}{137 143 740 467}

\clearpage

\figp{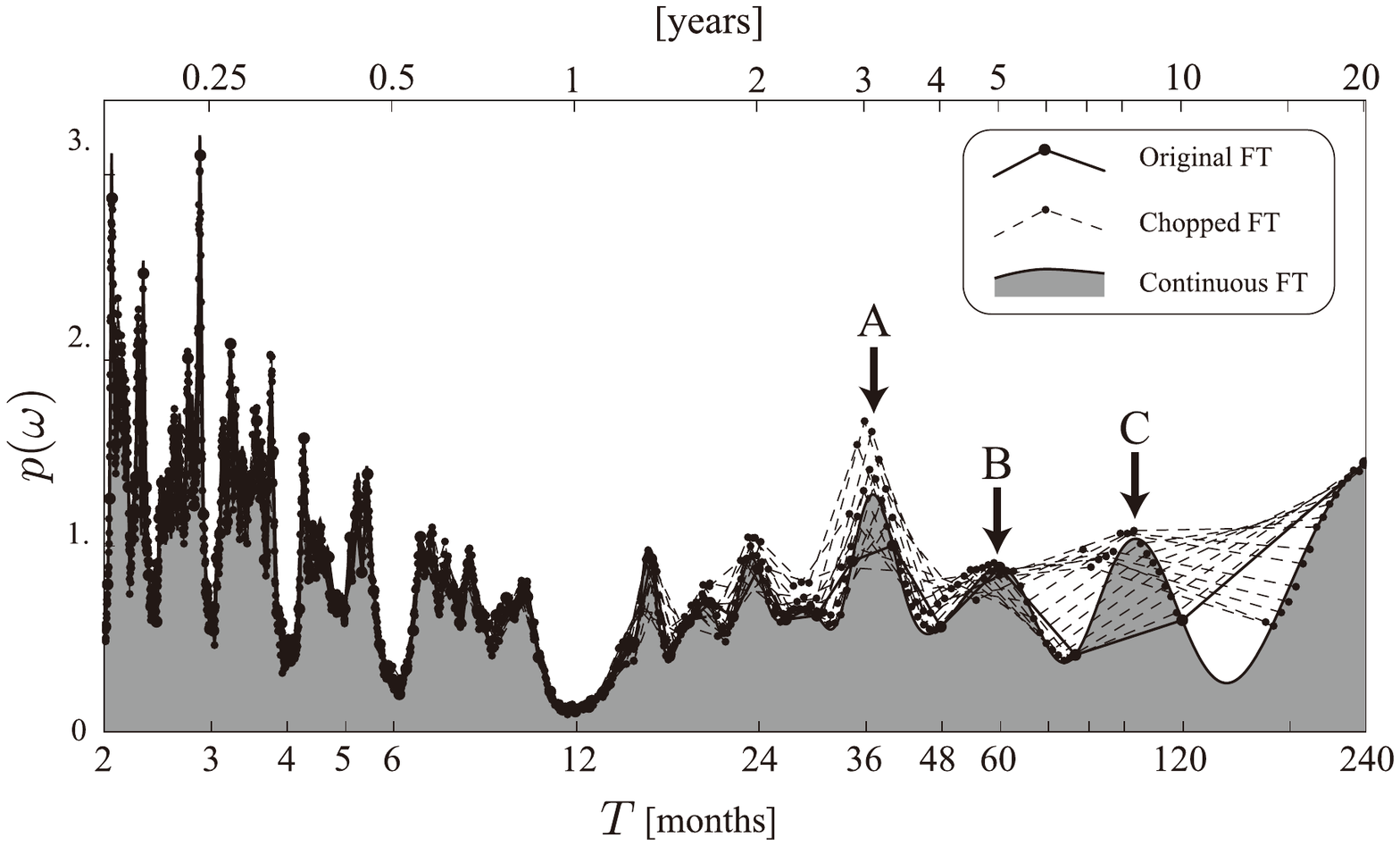}{Demonstration of robustness of the $T=60$ and $T=40$ cycles.
\protect\newline
Note: 
The big circles connected with solid lines (denoted as
``Original FT" on the top line of the legend) are the spectrum 
for the seasonally-adjusted data, the same as in Figure~2.
The small circles connected with dashed lines
(denoted as ``Chopped FT" on the second line of the legend)
are the results of the discrete Fourier transformation of
the first $S$ months as in the manner of Eq.~(\ref{defps}).
The dark shaded area shows
the result of the continuous Fourier transform
allowing $\omega$ to take any value in Eq.~(\ref{defps}).}{0.9}{29 267 550 580}

\clearpage

\figp{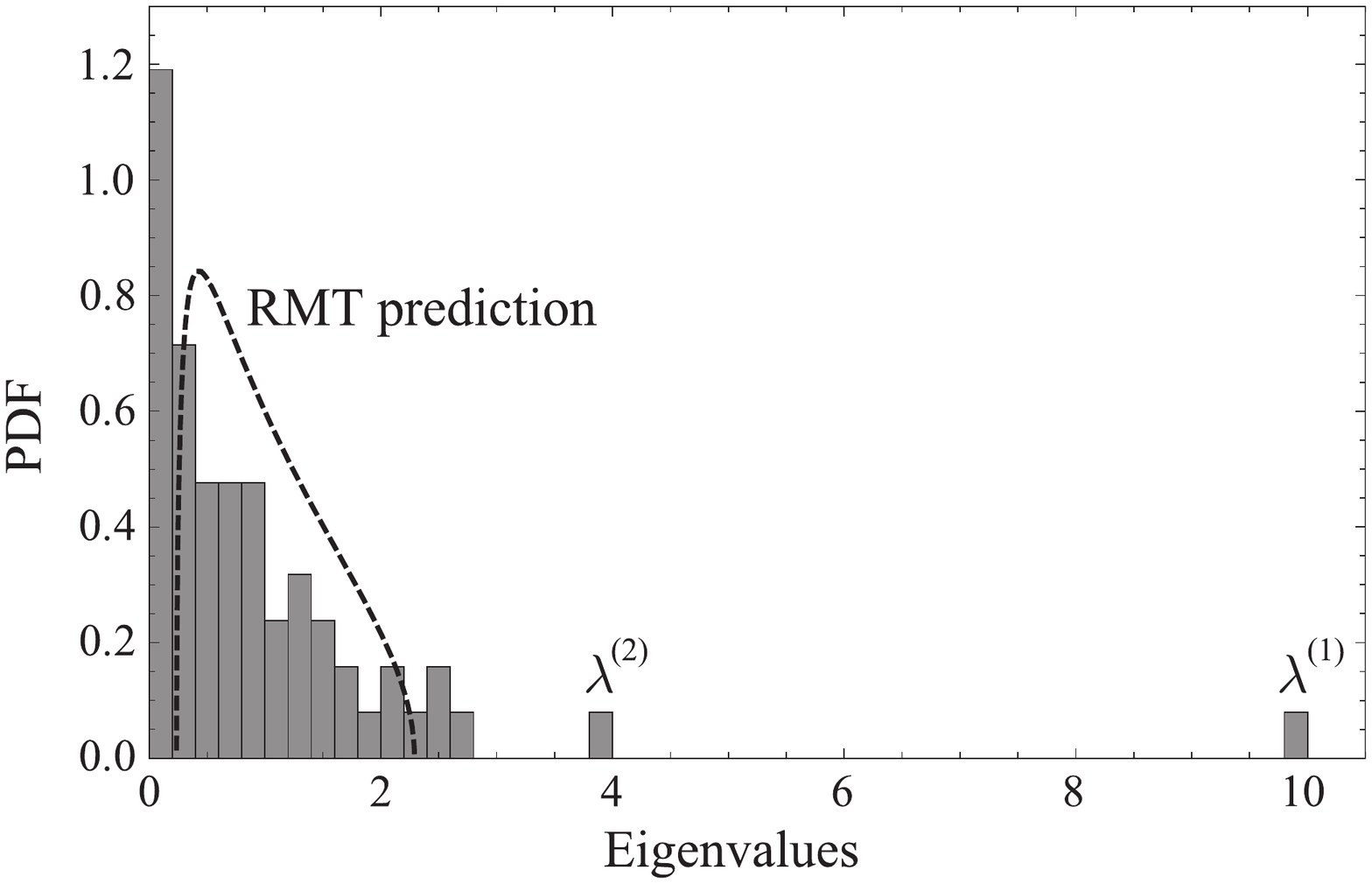}
{PDF of the eigenvalues (bars) and the RMT prediction $\rho(\lambda)$ (dashed curve).}{0.8}{127 105 670 454}

\clearpage
\figp{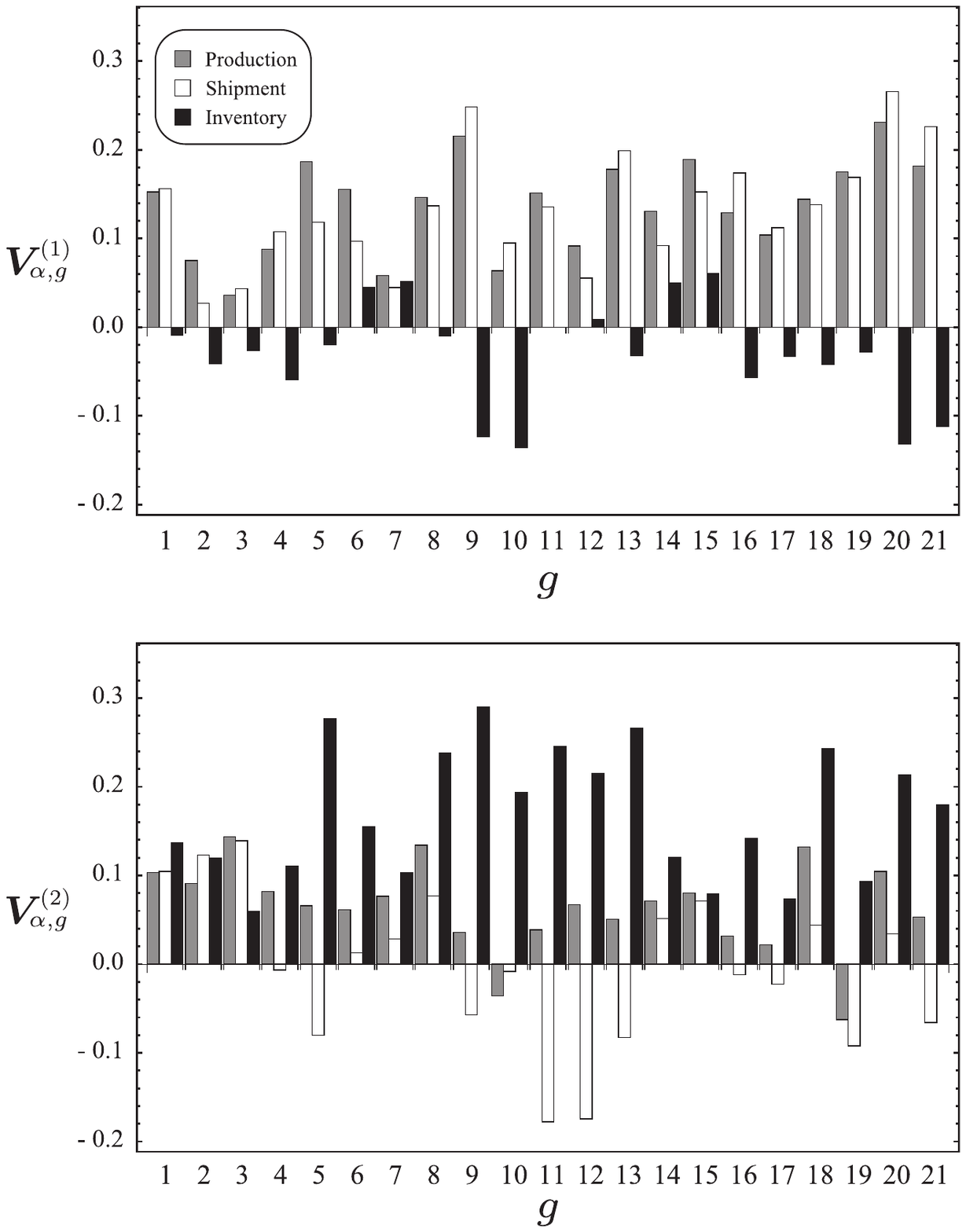}{The components of the two dominant eigenvectors, $\bm{V}^{(1,2)}$.
\protect\newline
Note: The horizontal axis shows the 21 goods $(g=1,2,\dots,21)$. 
The gray bars are the production ($\alpha=1$),
the white bars the shipments ($\alpha=2$),
and the black bars the inventory ($\alpha=3$).}{0.7}{152 52 528 534}

\clearpage
\figp{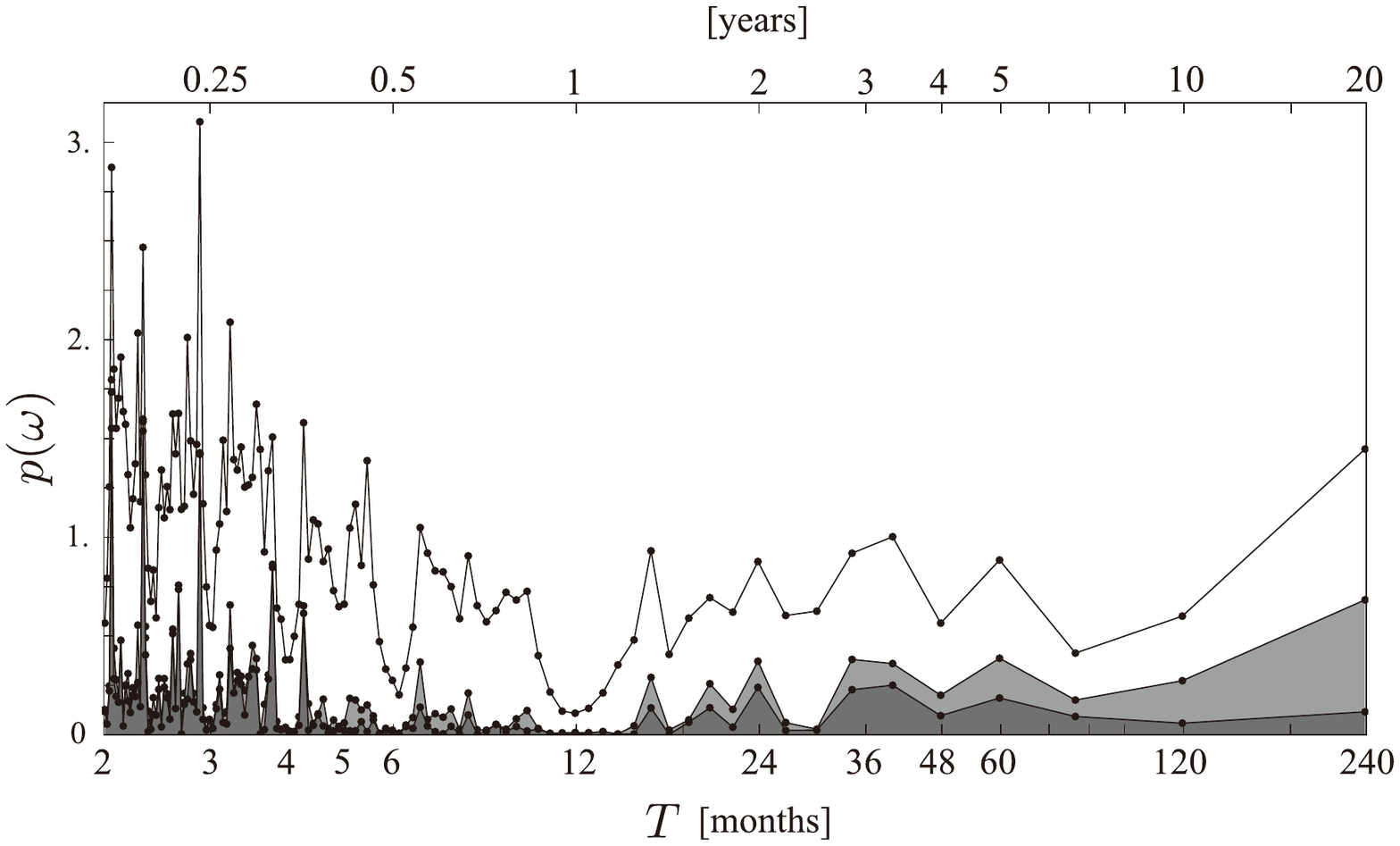}{Power spectrum for the
total and the two largest eigenmodes.
\protect\newline
Note:
The solid lines at the top  show the power spectrum, which is the same as
the solid lines with darker shading in Figure A.2.
(Note that the vertical axis is in linear scale in this plot.)
The darker shading show the power spectrum of the
first eigenmode, while the lighter shading show that of the
second eigenmode.}{0.8}{152 145 673 458}

\clearpage
\figp{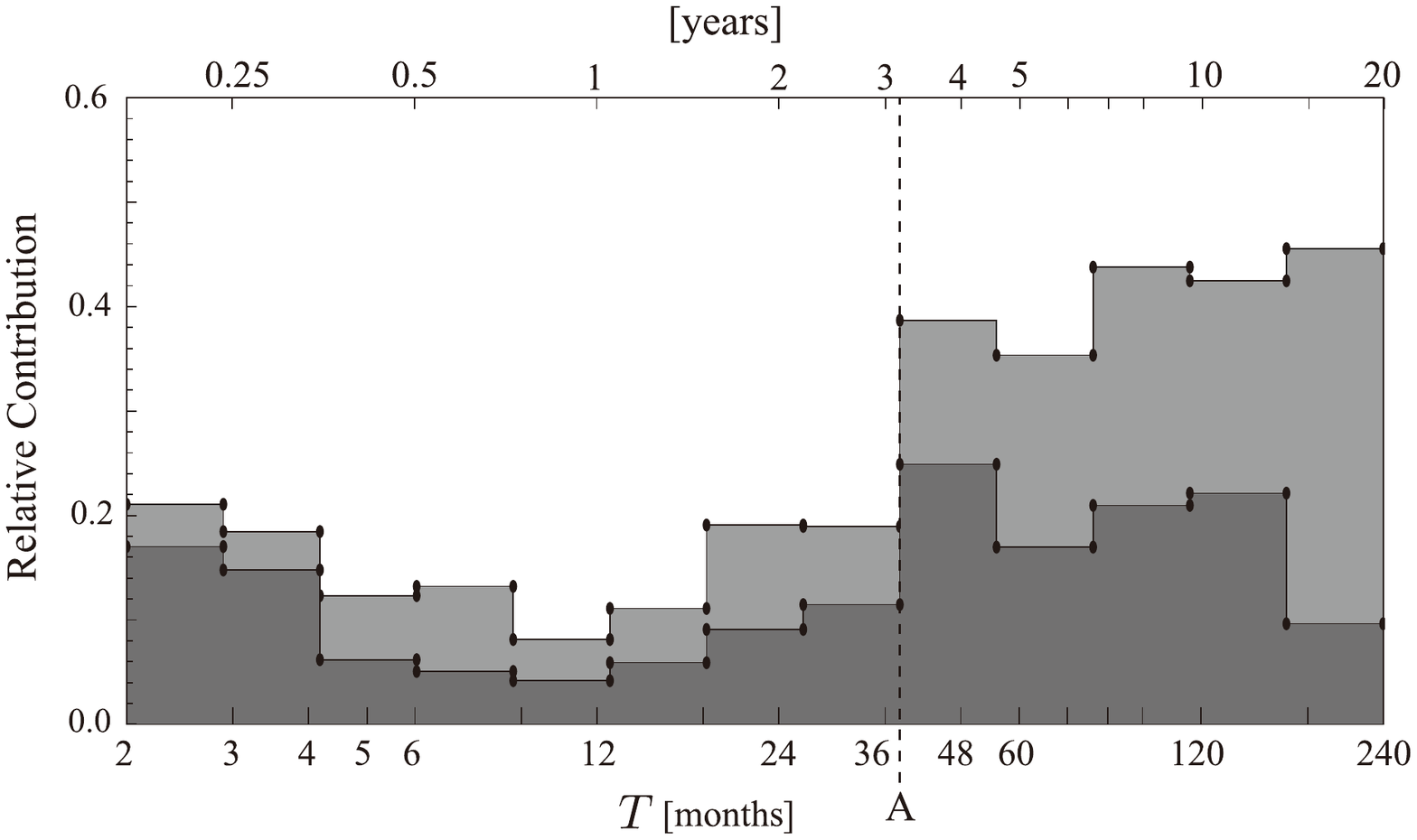}{The relative contribution of
the two largest eigenmodes to the power spectrum.
\protect\newline
Note:
The relative contributions are the values
averaged for each bin. The shading is the same as in Figure~6.
The point A on the horizontal is at $T\approx 38$ months.}{0.8}{144 159 674 471}

\clearpage
\figp{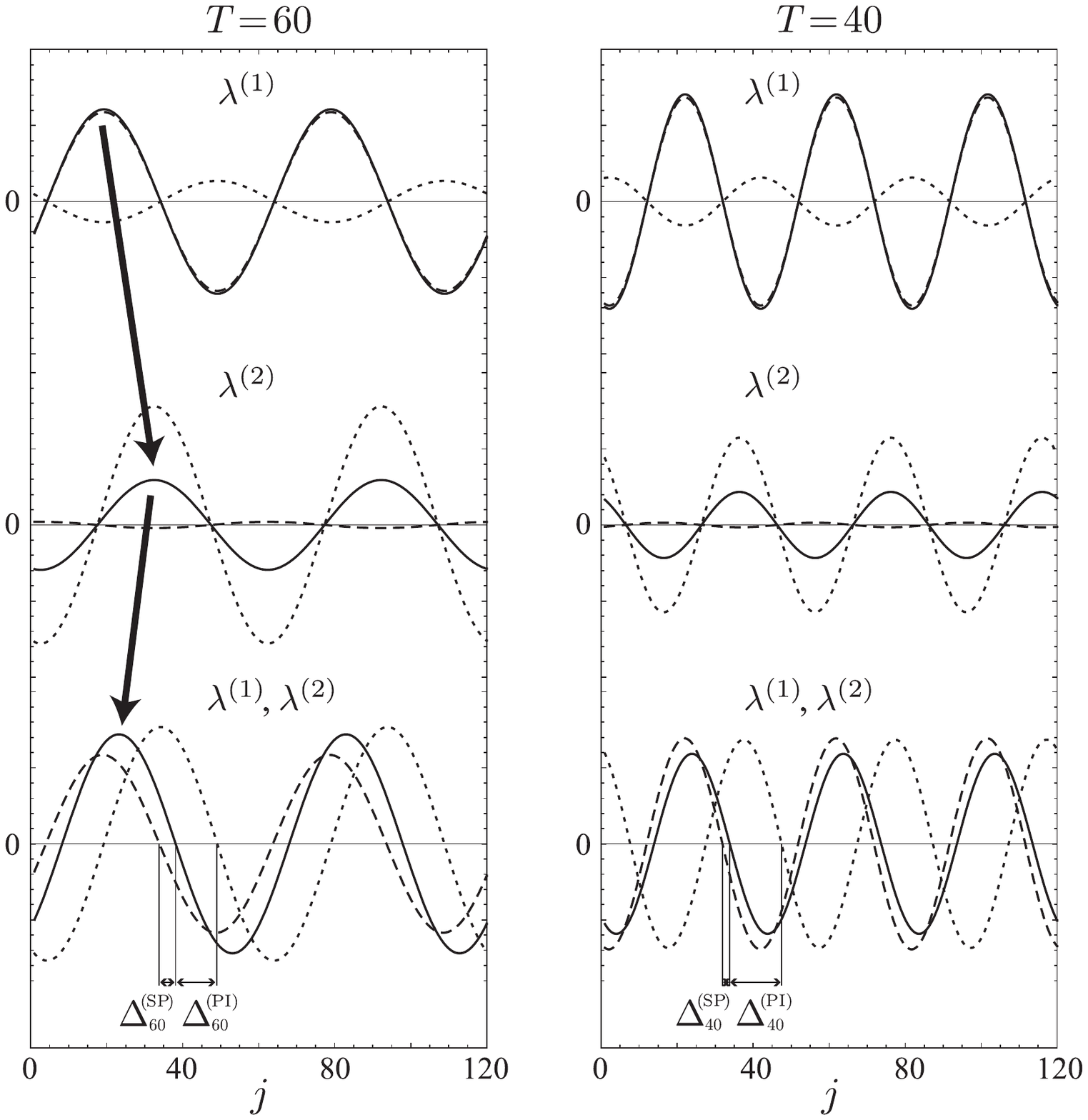}{The $T=60$ (left) and $T=40$ (right) oscillations
due to the first eigenvector (top), the second eigenvector (middle),
and their sum (bottom).
\protect\newline
Note: The solid line shows the production averaged over 21 goods,
the dashed line the average shipments, and the dotted line the average inventory.}{0.8}{83 81 561 573}

\clearpage
\figp{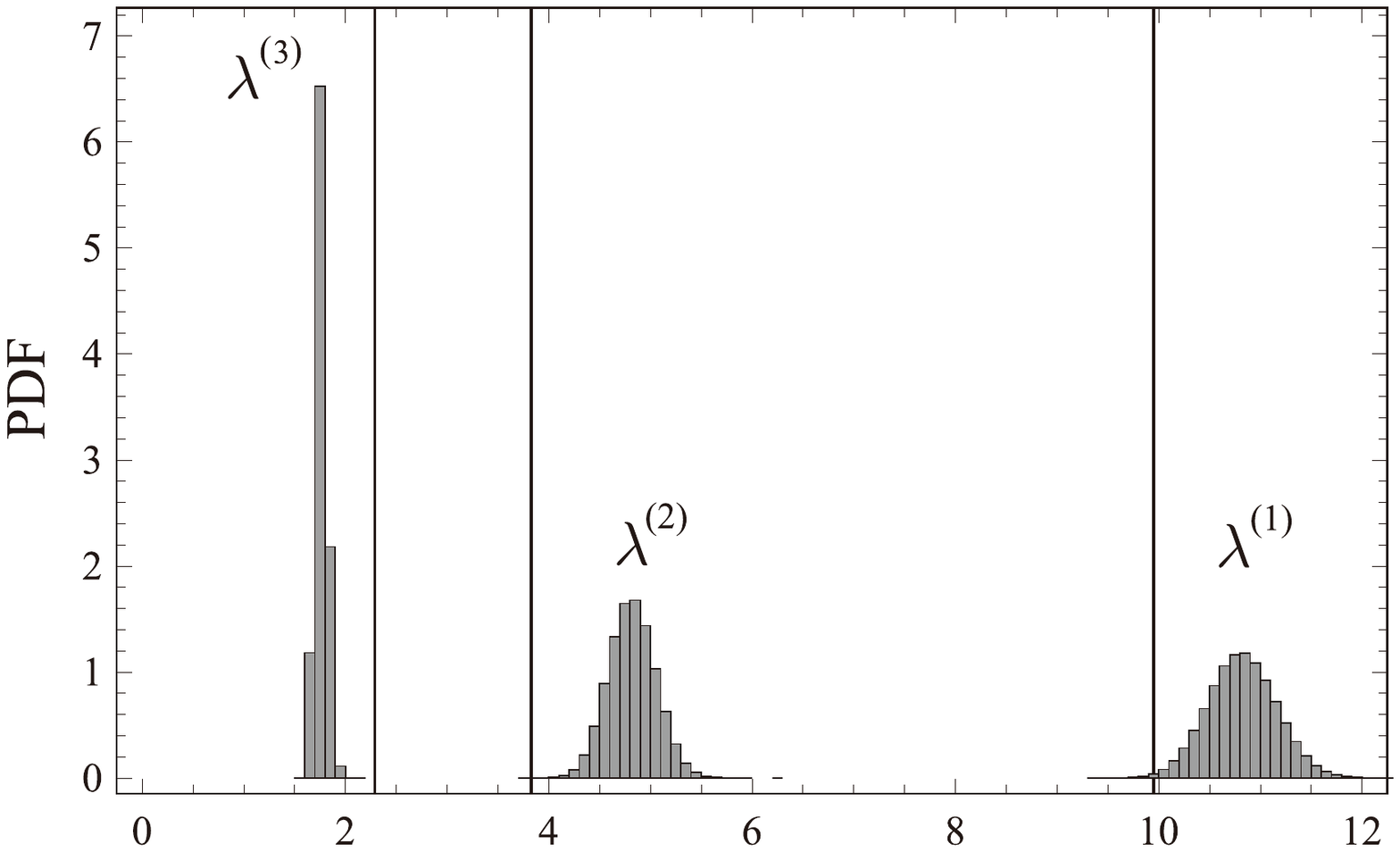}{PDF of the three largest eigenvalues obtained by the simulation. 
\protect\newline
Note: The three vertical lines show the RMT boundary,
the value of the true $\lambda^{(2)}$, and the value of the true $\lambda^{(1)}$
in the ascending order.}{0.7}{164 130 650 424}

\clearpage
\figp{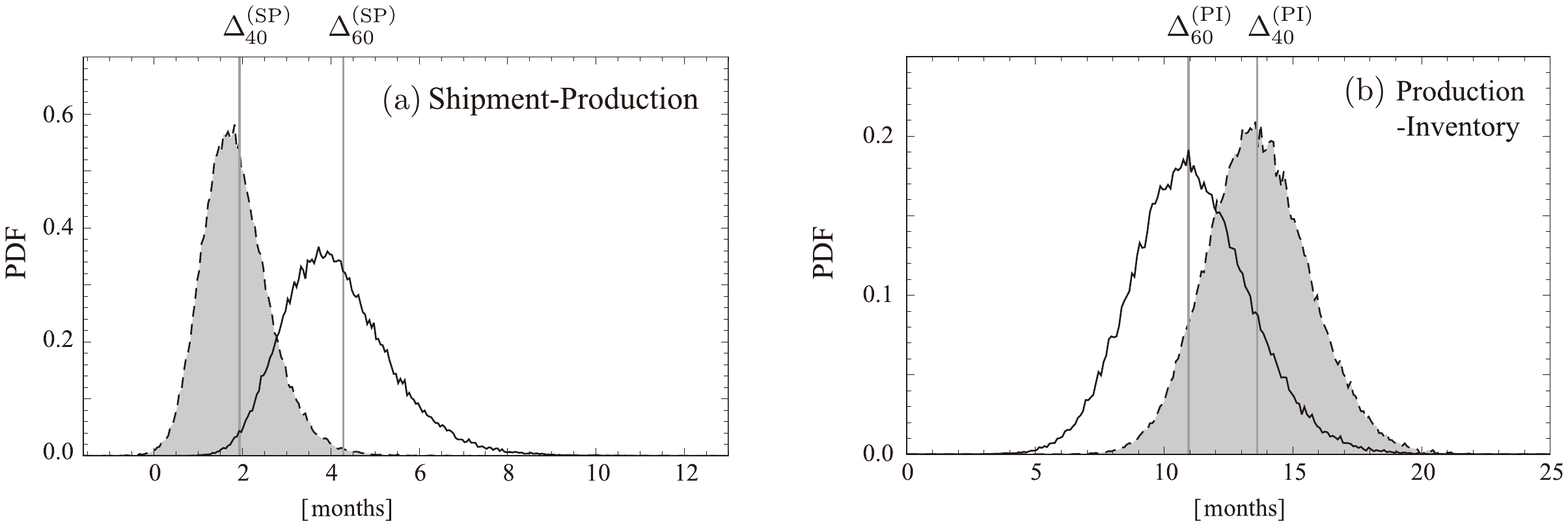}{PDF of  the time delays obtained by the simulation of $10^5$ times.
\protect\newline
Note: (a) $\Delta^{\rm (SP)}_{60}$ (solid lines) and
$\Delta^{\rm (SP)}_{40}$ (dashed lines with gray shading).
(b) $\Delta^{\rm (PI)}_{60}$ (solid lines) and
$\Delta^{\rm (PI)}_{40}$ (dashed lines with gray shading).
The vertical line in each plot shows the observed values.
The fact that the simulation results are distributed around the observed values
is an evidence that the observed values are independent from the noise part and 
are statistically significant. The statistical measures obtained in 
this simulation are given in Table~\ref{tab:suchi}.}{1}{62 180 751 408}

\clearpage

\figp{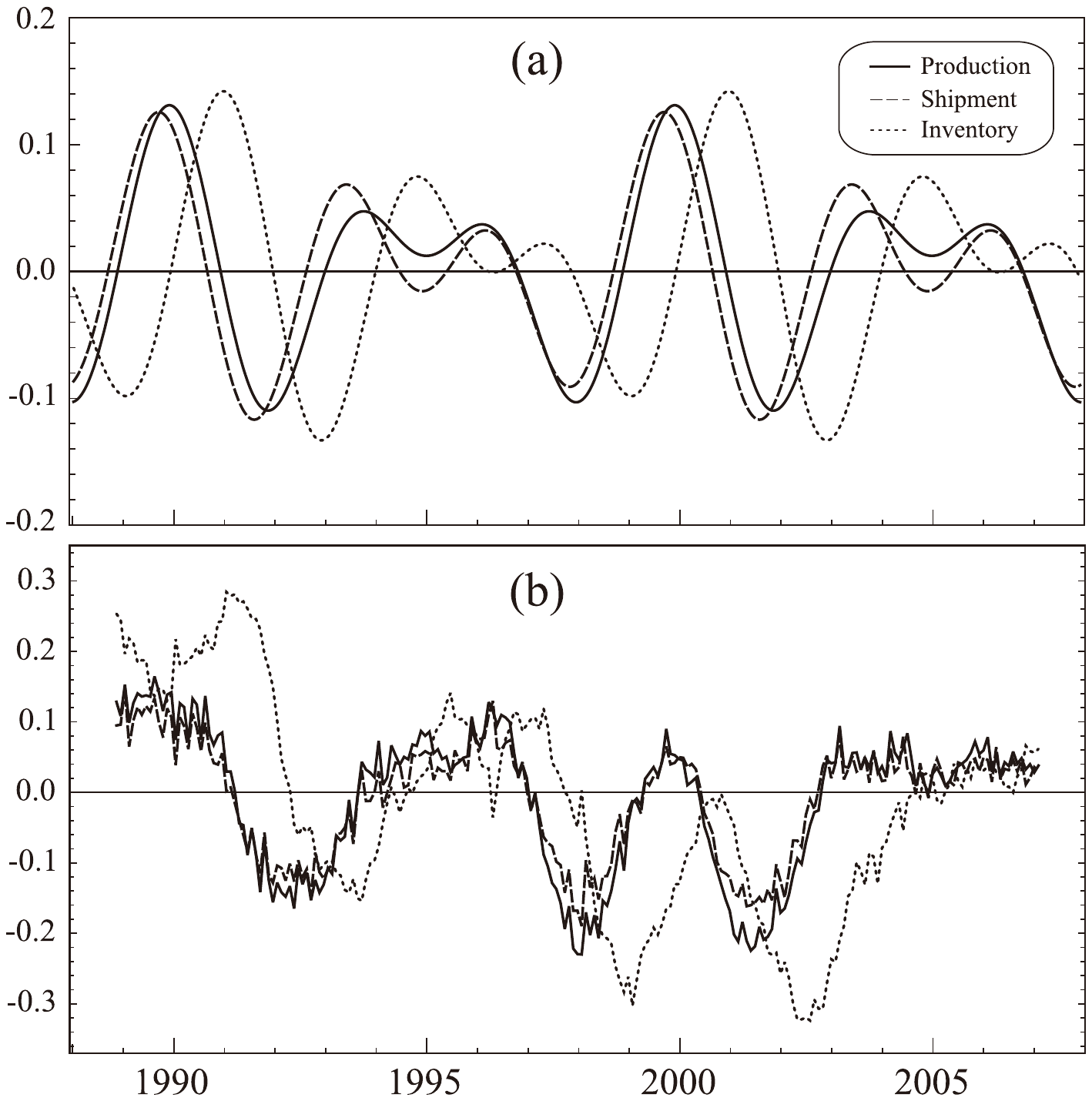}{Comparison between the extracted business cycles and the original data.
\protect\newline
Note: (a) Sum of the $T=60$, $40$ components of the first and the second eigenvectors
for production, shipments and inventory, each of which is averaged over 21 goods. 
(b) The original data shown in the bottom panel are smoothed out by taking one-year simple moving average.
We observe that the systematic ``business cycles'' are well extracted from the original data.
(Since the Fourier components of $T=120$ and $240$ are NOT included in the plot (a),
that part of the the behavior of the real data shown in (b) is not reproduced in the plot (a).)}
{0.7}{120 241 524 648}

\clearpage
\figp{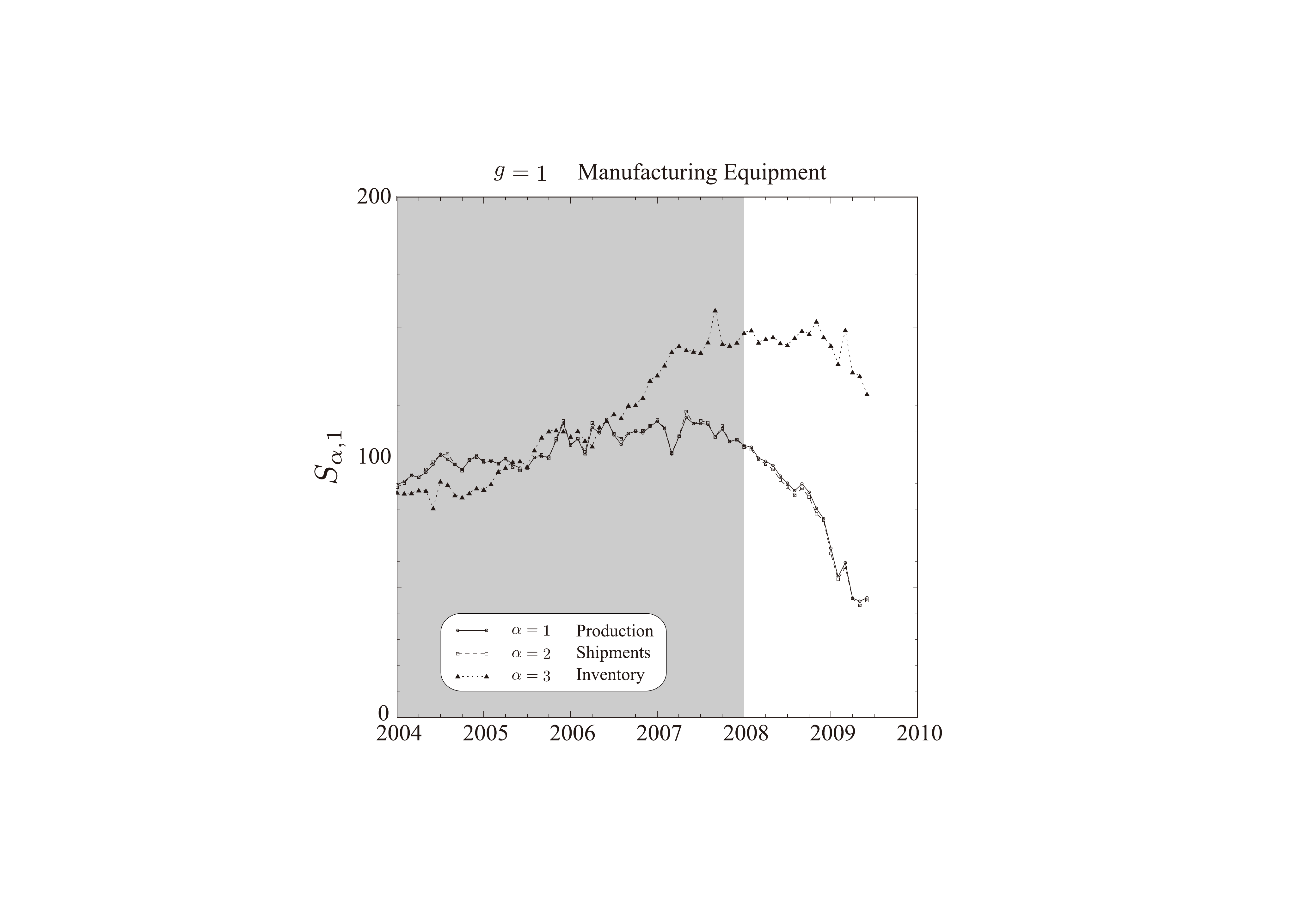}{The 2008--09 recession.
\protect\newline
Note: The IIP data for $g=1$ (Manufacturing Equipment)
shown in Figure~\ref{fig:data_g1new} is extended
to encompass the 2008--09 economic crisis. 
The gray shaded zone depicts the in-sample period 
used in the analyses in Sections 3--6.
Production and shipments almost overlap with each other
and are hard to distinguish.}{0.7}{286 166 858 693}

\clearpage
\figp{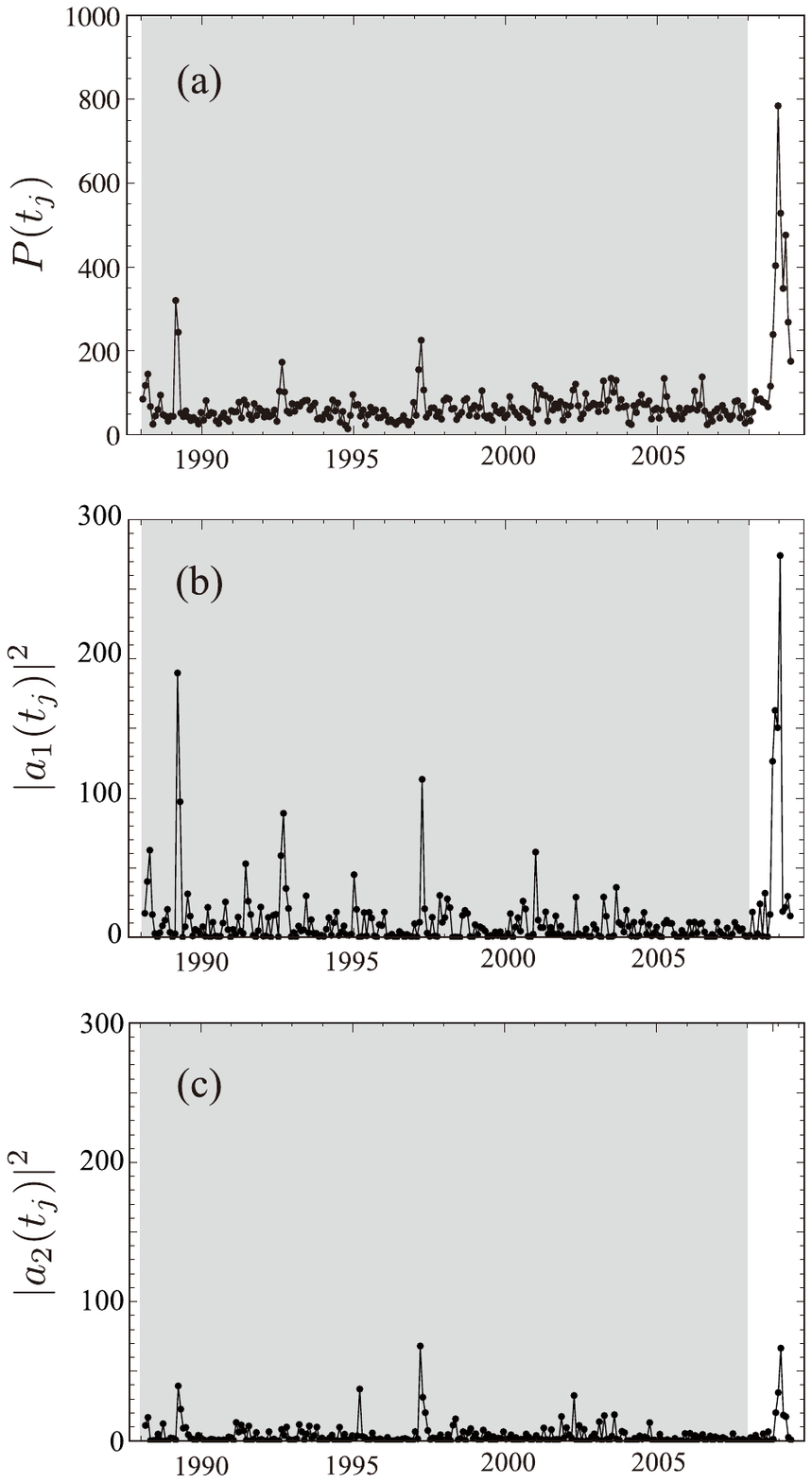}{Decomposition of volatility of the extended IIP data.
\protect\newline
Note: (a) The total volatility. 
(b) The partial contribution of the first eigenmode to the total volatility.
(c) The partial contribution of the second eigenmode. 
The gray shaded zone depicts the in-sample period used for determining 
the dominant modes.}{0.6}{155 237 423 730}

\clearpage
\figp{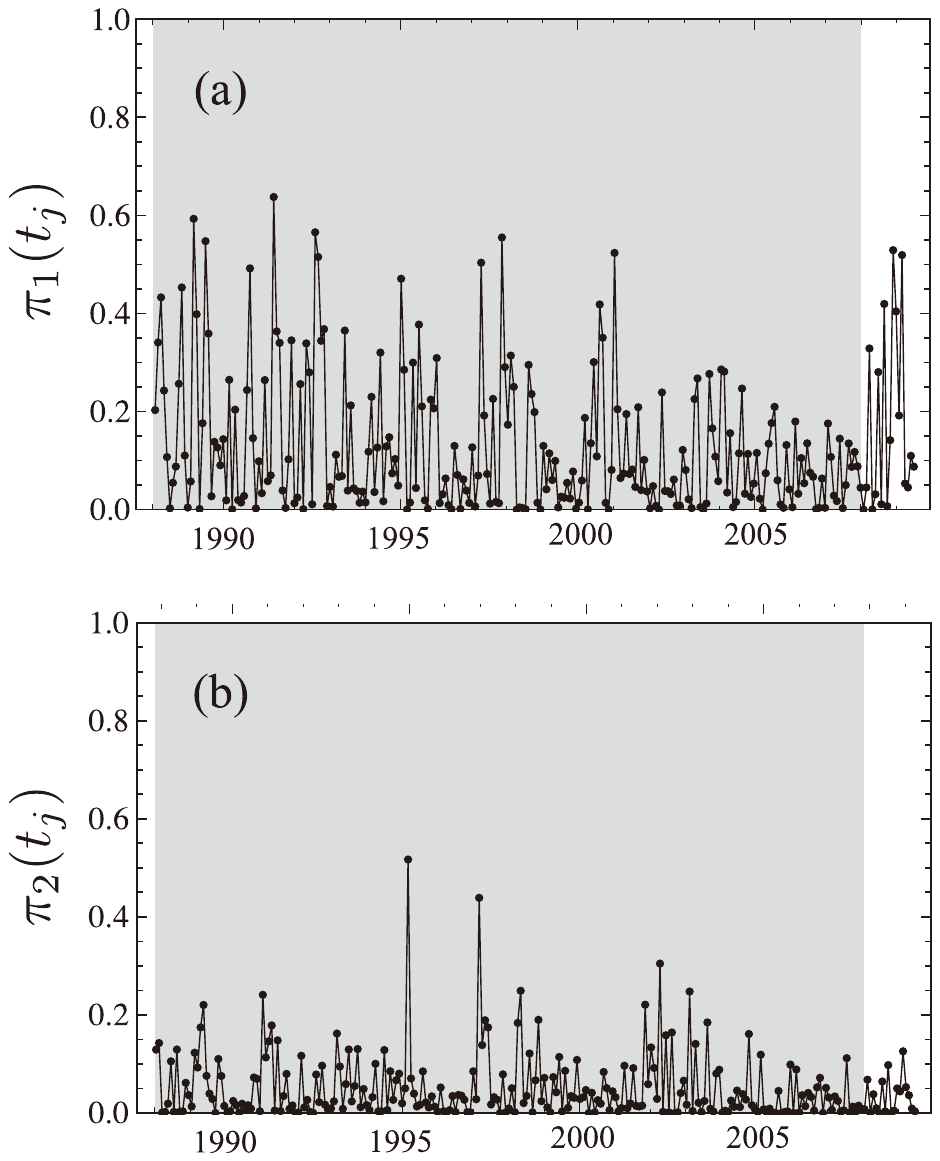}{Relative contributions of (a) the largest mode and 
(b) the second largest mode.
\protect\newline
Note: Panels (a) and (b) correspond to panels (b) and (c) in Figure~\ref{fig:crisis_abs}, respectively. }{0.65}{161 309 427 638}

\clearpage
\figp{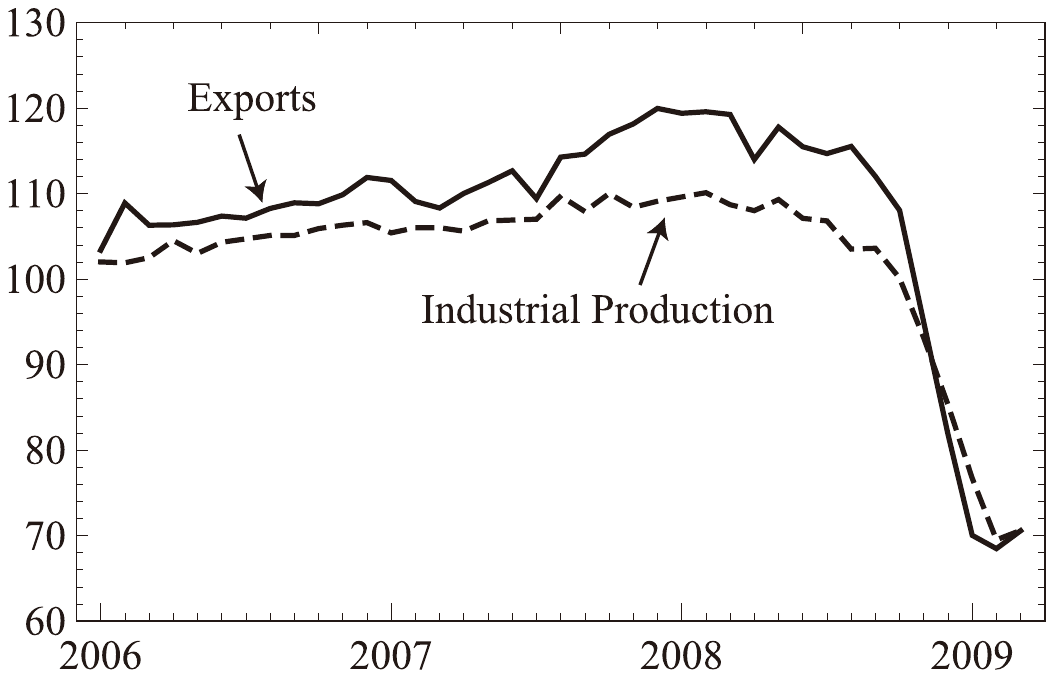}{Indices of exports (solid line) and the Industrial Production (dashed line),
normalized to 100 for 2005.}{0.7}{270 198 570 395}

\clearpage
\figp{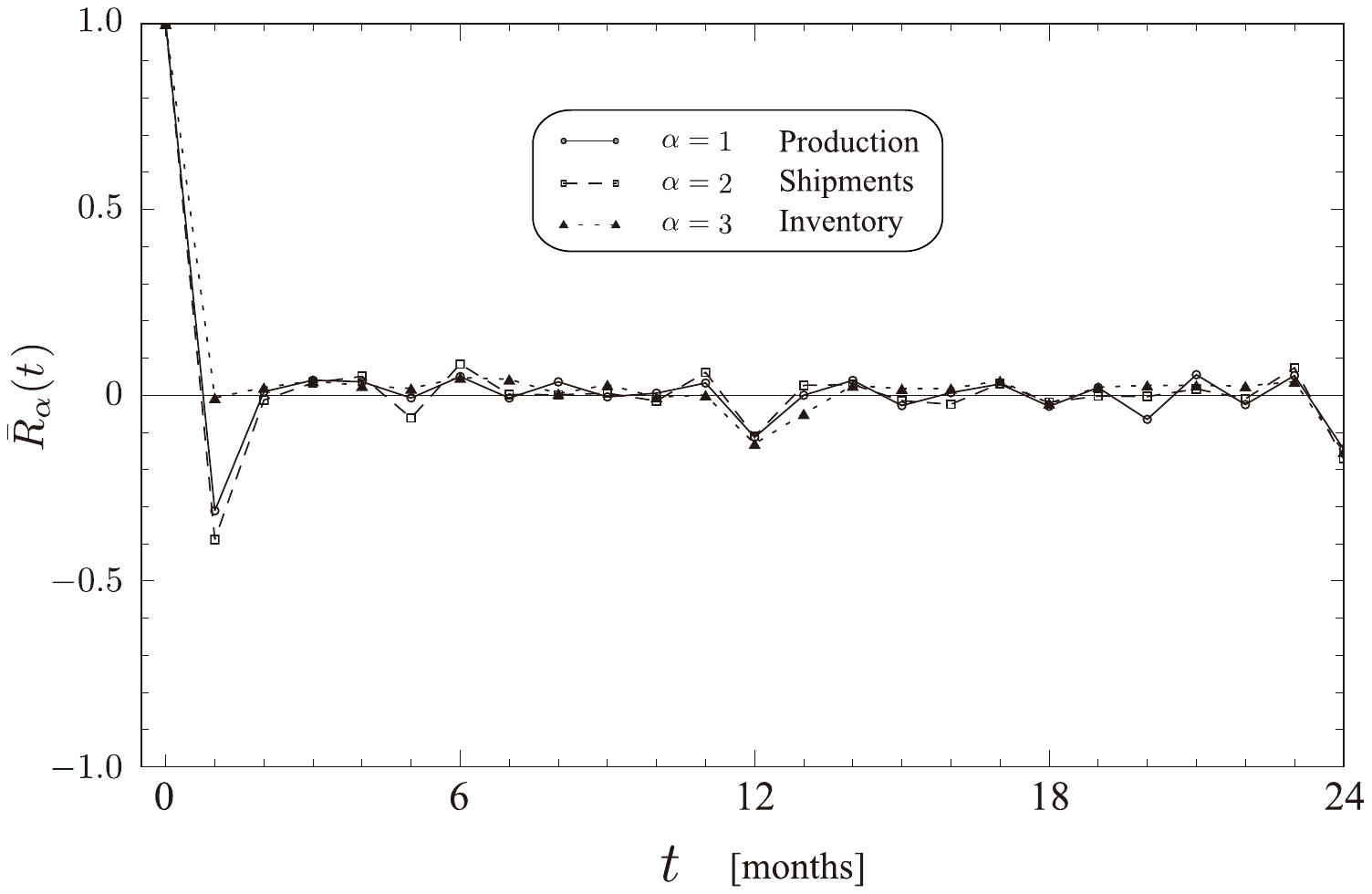}{Autocorrelation of the IIP growth rate, $\bar{R}_\alpha$,
defined by Eqs.(\ref{auto1}) and (\ref{auto2}).}{0.7}{74 269 595 546}

\clearpage
\figp{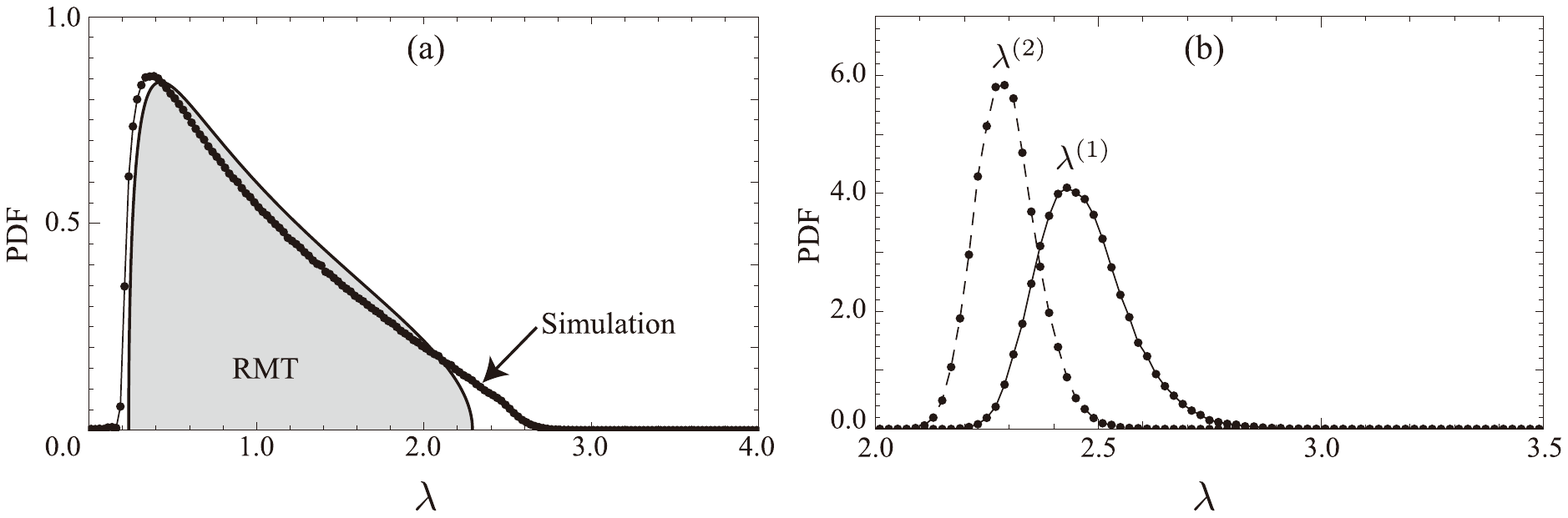}{Results of the random-ordering simulation.
\protect\newline
Note:
Because of the fact that the IIP data is of finite size and have nontrivial
autocorrelation, the distribution of eigenvalues could differ from that of
RMT prediction, even if the IIP data lacks any comovements. 
The plot (a) shows that the PDF of the eigenvalues $\lambda$,
where the gray-shaded area is the RMT result and the dots connected by lines
show the result of the simulation.
The plot (b) shows the PDF of the first eigenvalue $\lambda^{(1)}$
and the second eigenvalue $\lambda^{(2)}$ obtained by simulation.
From these, we conclude that $\lambda^{(1)}$ and $\lambda^{(2)}$ found in the actual data
are in fact principal factors that correspond to comovements.}{1}{49 320 586 495}

\clearpage
\figp{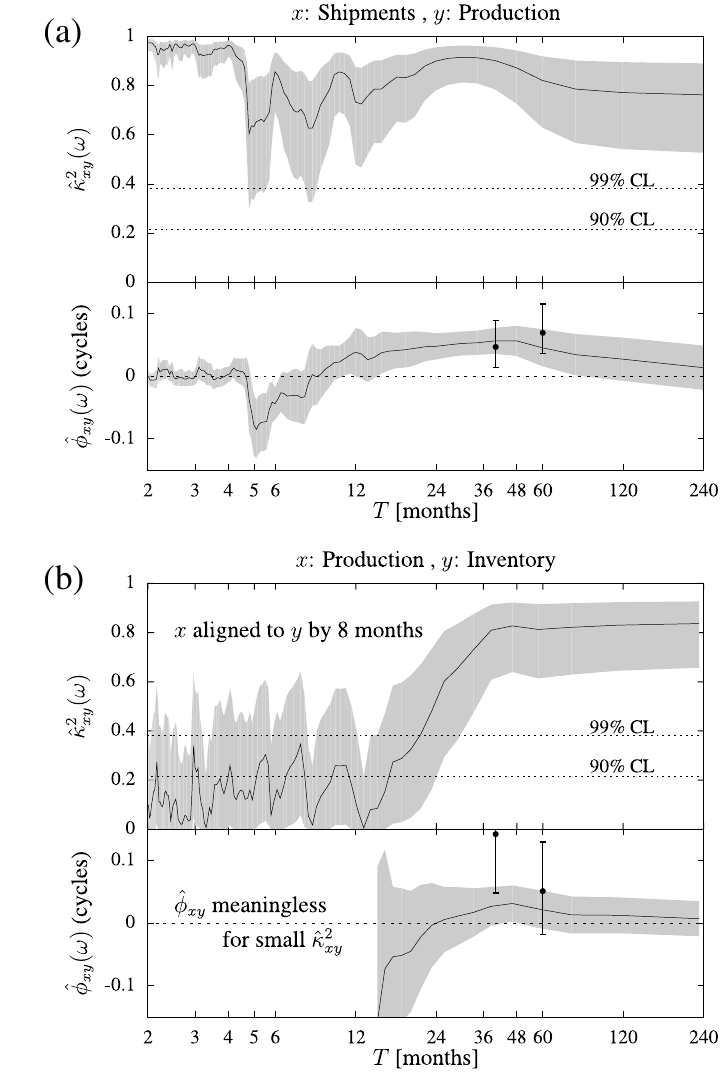}{Cross-spectrum for the sum of two dominant eigenmodes;
(a) for production and shipments and
(b) for production and inventory.
\protect\newline
Note: In each plot, $\hat{\kappa}_{xy}^2(\omega)$ is the squared
coherency, and $\hat{\phi}_{xy}(\omega)$ is the phase of the
cross-spectrum.  In both panels, the dotted lines are 90\% and 95\%
confidence levels for $\hat{\kappa}_{xy}^2(\omega)$ obtained by the
null hypothesis of $\hat{\kappa}_{xy}^2(\omega)=0$. The gray areas
represent 95\% confidence intervals in each plot.}{0.7}{0 0 208 312}

\clearpage
\figp{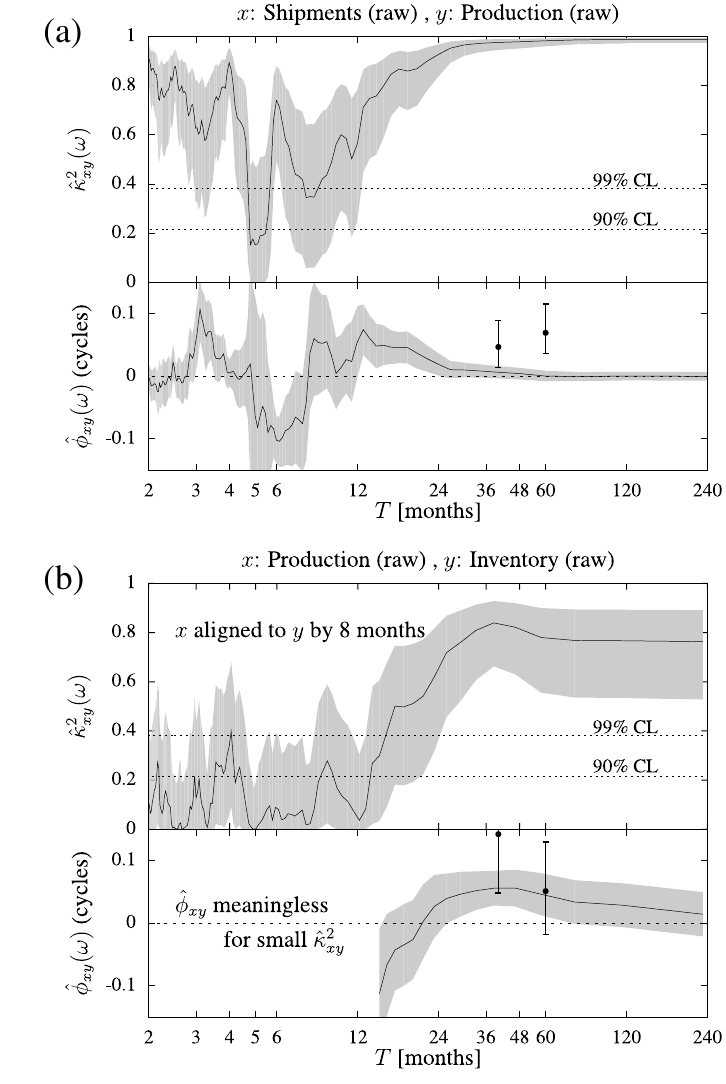}{Cross-spectrum for the original time-series;
(a) for production and shipments and
(b) for production and inventory.
\protect\newline
Note: $\hat{\phi}_{xy}$ for the production and shipments is not
significantly different from zero at $T=40,60$ in the plot of of upper
panel. On the other hand, in the lower panel, production leads
inventory by a significant difference in the phase.}{0.7}{0 0 209 312}

\clearpage

\renewcommand{\thetable}{\arabic{table}}

\setlength{\doublerulesep}{4pt} 
\renewcommand{\arraystretch}{1.1}
\begin{table}\small
\centering
\caption{Classifications of 21 goods in IIP.}
\label{tab:21}
\vspace{2mm}
\begin{tabular}{ccl}
\hline
\multicolumn{3}{>{\columncolor[gray]{0.6}}c}{\sf Final Demand Goods}\\ \hline
\multicolumn{3}{c}{\sf Investment Goods}\\
\hline
Capital Goods &1& Manufacturing Equipment \\
&2& Electricity\\
&3&Communication and Broadcasting\\
&4&Agriculture\\
&5& Construction\\
&6&Transport\\
&7&Offices\\
&8&Other Capital Goods\\
\hline
Construction Goods&9&Construction\\
&10&Engineering\\
\hline
\multicolumn{3}{c}{\sf Consumer Goods}\\
\hline
Durable Consumer&11& House Work\\
Goods&12& Heating/Cooling Equipment\\
&13&Furniture \& Furnishings\\
&14&Education \& Amusement\\
&15&Motor Vehicles\\
\hline
Non-durable &16& House Work\\
Consumer Goods&17& Education \& Amusement\\
&18& Clothing \& Footwear\\
&19& Food \& Beverage\\
\hline\hline
\multicolumn{3}{>{\columncolor[gray]{0.6}}c}{\sf Producer Goods}\\ 
\hline
&20& Mining \& Manufacturing \\
&21& Others\\
\hline
\end{tabular}\\
\vspace{4pt}
\centerline{Note: The center column is the identification number, $g$.}
\end{table}

\clearpage

\begin{table}
\begin{center}
\caption{Simulation results on the time delays
in comparison with estimates based on the original data.
\protect\newline
Note: $\sigma$ is the standard deviation, and ``95\% CI'' is the Confidence Interval
at the 95\% of the distribution of the simulation results shown in Figure~\ref{fig:simulation2}.}
\label{tab:suchi}
\vspace{2mm}
\begin{tabular}{r|r|rcc}
\hline
& Original data & Average & $\sigma$ & 95\% CI\cr
\hline
\rule[0pt]{0pt}{13.5pt}
$\Delta^{\rm (SP)}_{60}$ & 4.28 & 4.16 & 1.22 & $[\, 2.17, 6.91\, ]$\cr
$\Delta^{\rm (SP)}_{40}$ & 1.93 & 1.87 & 0.77 & $[\, 0.57, 3.57\, ]$\cr
\hline
\rule[0pt]{0pt}{13.5pt}
$\Delta^{\rm (PI)}_{60}$ & 10.95 & 11.07 & 2.25 & $[\, 6.92, 15.77\, ]$\cr
$\Delta^{\rm (PI)}_{40}$ & 13.62 & 13.67 & 2.00 & $[\, 9.94, 17.78\, ]$\cr
\hline
\end{tabular}
\end{center}
\end{table}



\clearpage

\section*{Footnotes}
\begin{enumerate}
\item
\citet{shapiro1988sbc} consider labor supply shocks as well as technology shocks, and obtain the ``surprising'' result that the labor supply shocks account for at least 40 percent of output variation at all horizons. However, \citet{hall1988csw} shows that this finding is neither surprising nor plausible. Specifically, as Hall criticizes, \citet{shapiro1988sbc} make the implausible identifying assumption that all the contemporaneous comovements of output and work hours be attributed to shifts in labor supply, and rule out the alternative assumption, perhaps more plausible to many, that the moving force operates from output to hours in the short-run.
This paper indeed demonstrates that output is basically determined by real demand in business cycles.
\item
This reduces the number of independent real components in the complex Fourier
coefficients $\tw$'s from $2N'-1$ down to $N'$, guaranteeing that 
the Fourier coefficients $\tw$'s are the faithful representation of the original series $w$.
\item
The seasonal adjustment is done by the U.S. Census Bureau's X-12 ARIMA method. The details are given in the resources available from \citet{iip}.
\item
Note that the vertical scale in Figure~\ref{fig:powerspectrum} is logarithmic.
\item
There are, in fact, slight dips at those frequencies, which
means that the seasonal adjustments are somewhat 
excessively done. However, we 
have also looked at the auto-correlation of the data
and have found that the adjusted data is much better than that of
the non-adjusted data.
The relevant results are available from authors upon request.
\item
These variables must be normally distributed and uncorrelated.
\item
We note that our extraction of the first two dominant eigenmodes is
essential for obtaining the lead-lag relationship between shipments and production.
In fact, if we use all the eigenvectors, which is equivalent to the usual Fourier
decomposition, we obtain $-0.10$ months instead of 4.28 months in Eq.(\ref{dsp60})
and 0.34 months instead of 1.93 months in Eq.(\ref{dsp40}).
They are much smaller than our time-increment of 1 month,
and therefore, are too small to be taken as being significantly different from zero.
\end{enumerate}

\end{document}